\documentclass[onecolumn,groupedaddress]{revtex4}
\usepackage{graphicx}
\usepackage{epstopdf}
\usepackage{latexsym}
\usepackage{amssymb}
\usepackage{amsfonts} 
\usepackage{txfonts}
\usepackage{pxfonts}
\usepackage{color}
\usepackage[usenames,dvipsnames]{xcolor}
\usepackage{pdfpages}

\newcommand{\dd}{\textrm{d}}

\usepackage[super]{nth}
\usepackage{subfigure}

\begin{document}

\title{Perturbing the catenoid: stability and mechanical properties of non-axisymmetric minimal surfaces$^\dag$}

\author{Friedrich Walzel,\textit{$^{a}$\ddag} Alice Requier,\textit{$^{a,b}$\ddag} Kevin Boschi,\textit{$^{a}$} Jean Farago,\textit{$^{a}$} Philippe Fuchs,\textit{$^{a}$} Fabrice Thalmann, Wiebke Drenckhan,\textit{$^{a}$} Pierre Muller,\textit{$^{a}$} and Thierry Charitat,$^{\ast}$\textit{$^{a}$}}
\affiliation{Institut Charles Sadron, Universit\'e de Strasbourg, CNRS, 23 rue du Loess, BP 84047 67034 Strasbourg Cedex 2, France}
\date{\today}

\begin{abstract}
Minimal surface problems arise naturally in many soft matter systems whose free energies are dominated by surface or interface energies. Of particular interest are the shapes, stability and mechanical stresses of minimal surfaces spanning specific geometric boundaries. The "catenoid" is the best-known example where an analytical solution is known which describes the form and stability of a minimal surface held between two parallel, concentric circular frames. Here we extend this problem to non-axisymmetric, parallel frame shapes of different orientations, by developing a perturbation approach around the known catenoid solution. We show that the predictions of the perturbation theory are in good agreement with experiments on soap films and finite element simulations (Surface Evolver). Combining theory, experiment and simulation, we analyse in depth how the shapes, stability and mechanical properties of the minimal surfaces depend on the type and orientation of elliptical and three-leaf clover shaped frames. In the limit of perfectly aligned non-axisymmetric frames, our predictions show excellent agreement with a recent theory established by Alimov et al (M. M. Alimov, A. V. Bazilevsky and K. G. Kornev, Physics of Fluids, 2021, 33, 052104). Moreover, we put in evidence the intriguing capacity of minimal surfaces between non-axisymmetric frames to transmit a mechanical torque despite being completely liquid. These forces could be interesting to exploit for mechanical self-assembly of soft matter systems or as highly sensitive force captors.
\end{abstract}

\maketitle

\footnotetext{\textit{$^{b}$~Present Address Laboratoire de Physique des Solides, Universit\'e Paris-Saclay, UMR CNRS 8502, B\^atiment 510, 91405 Orsay Cedex, France}}

\footnotetext{\ddag~These authors contributed equally to this work}

\section{Introduction}

Minimal surfaces describe shapes of arbitrarily complex geometry which are characterised by the fact that their surface has a minimal area fixed by a set of boundary conditions. Minimal surfaces are fascinating mathematical objects introduced by the pioneering work of \cite{euler1744methodus}, Lagrange \cite{lagrange1760essai} and Plateau \cite{plateau1873statique}. Today, they are related to different mathematical fields such as calculus of variations, partial differential equations, differential geometry and topology, and complex analysis via the Weierstrass representation \cite{nitsche1989lectures,Meeks2012ASO}. Minimal surfaces have also served as models for numerous applications. Examples include architecture \cite{Emmer2013} or the development of materials combining antagonistic properties such as good mechanical rigidity and high electrical/thermal transport capacities \cite{torquato(PRL2002),Zhou(2007),CHEN2010806}. We can also mention their use as scaffold for tissue engineering \cite{KAPFER20116875}.

The relevance of minimal surfaces in different fields leads to different (yet equivalent) mathematical definitions \cite{Meeks2012ASO}. The property of having a zero mean curvature at all points of the surface is particularly remarkable \cite{meusnier1785memoire}. For physicists, the property of minimising the area of the surface is even more important, since it makes it possible to make the connection with physical problems where surface area can be associated with the energy of a system which needs to be minimised. Soap films are amongst the most popular examples \cite{plateau1873statique,courant1940}.

\begin{figure}[h]
\includegraphics[width=0.75\textwidth]{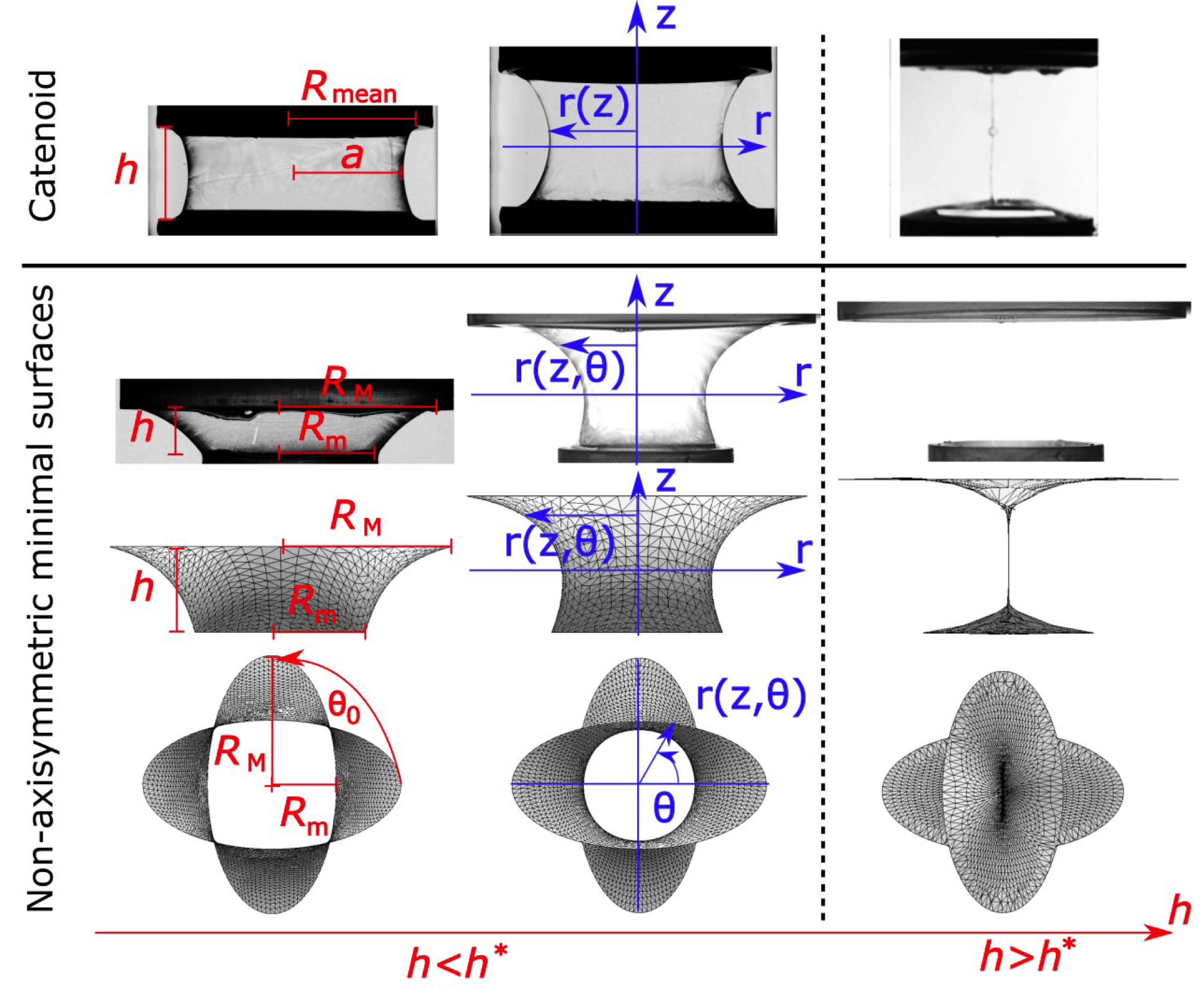}
\caption{Experimental (\nth{1} and \nth{2} row) and numerical (\nth{3} and \nth{4} row) shapes obtained for minimal surfaces with axisymmetric circular frames (\nth{1} row) and elliptic boundary conditions (eccentricity $e=0.866$) and for an angle $\theta_0=90^\circ$ between the main axes (\nth{2} to \nth{4} row). The distance between the upper and lower frame is increasing from left to right. The rightmost images correspond to transient since $h$ is already larger than the critical distance.}
\label{figure1} 
\end{figure}


The energy $\cal E$  of soap films with uniform surface tension $\gamma$ is given by
\begin{equation}
\label{eq:E_gamma_A}
{\cal E} = 2 \gamma \iint_{\cal A} \dd A = 2\gamma A,
\end{equation}
where integration is done over the surface $\cal A$, and $A$ is the total area of the film. The factor "2" results from the fact that a film has two interfaces. The equilibrium shape is then given by the minimisation of the total area $A$, which naturally leads to an easy realisation of minimal surfaces. 

A specific class of soap film problems was defined by Plateau. The so-called "Plateau problem" consists in studying the existence of minimal surfaces resting on given boundary conditions \cite{plateau1873statique,Plato,Courant97}. The catenoid is the classical example for this group of minimal surfaces. It consists of a minimal surface spanning two parallel circles of radius $R$ separated by a distance $h$ with the two centres of the circles lying on an axis orthogonal to the planes of each circle. 
Contrary to the majority of minimal surfaces, an analytical solution  is known for the catenoid \cite{euler1744methodus,meusnier1785memoire,Erle1970StabilityOI,toponogov2006differential}. It predicts the conditions for the existence of a solution and the exact shape of the minimal surface. The axisymmetric boundary conditions ensure that the surface is also axisymmetric. It can thus be described simply in cylindrical coordinates by a function giving the film radius $r_c$ depending on the vertical coordinate $z$ (see Fig. \ref{figure1})
in the range $-h/2$ to $h/2$
\begin{equation}
\label{eq:catenoid_shape}
r_c\left(z\right) = a_c\cosh{\left(\frac{z}{a_c}\right)}.
\end{equation}
The boundary conditions for the two frames are given by $r_c(z=\pm h/2)=R$. The smallest radius is found in the mid-plane ($z=0$) and is called the neck radius. Its value, $a_c$, is obtained with the equation 
\begin{equation}
\label{eq:neck_radius}
R = a_c\cosh{\left(\frac{h}{2a_c}\right)}.
\end{equation}
 This equation has two solutions for $h/R<\mu^\star$, one for $h/R=\mu^\star$ and no solution for $h/R>\mu^\star$ where $\mu^\star$ is the solution of the exact transcendental equation $(\mu^\star/2)\sinh((\mu^\star/2)\sqrt{1+4/{\mu^\star}^2})=1$ leading to $\mu^\star\approx 1.33$. In the following $h^\star$ will be the critical height defined by $h^\star=\mu^\star R$. If there are two solutions, one of them always corresponds to a maximum of the area $A$ and the other to a minimum. Experimentally, only the minimum is observable since the maximum is physically unstable for open systems (Erle {\it et al.} \cite{Erle1970StabilityOI} were able to observe the second solution in the case of closed systems). Close to $h^\star$, a small increase of the surface area leads to a destabilisation of the catenoid, which undergoes a topological instability leading to the so-called "Goldschmidt solution" \cite{sagan1992introduction} given by two planes parallel to the frames. 

The case of the catenoid has been much studied by physicists who have been particularly interested in the stability of the surface \cite{durand1981,amar1998stability,Jana2013}, in the collapse of the catenoid surface towards the Goldschmidt surface at the critical point \cite{cryer1992collapse,goldstein2021}, or in the asymmetrical catenoid supported by rings of different sizes \cite{Salkin}. 

\begin{figure}
    \centering
    \includegraphics[width=0.4\textwidth]{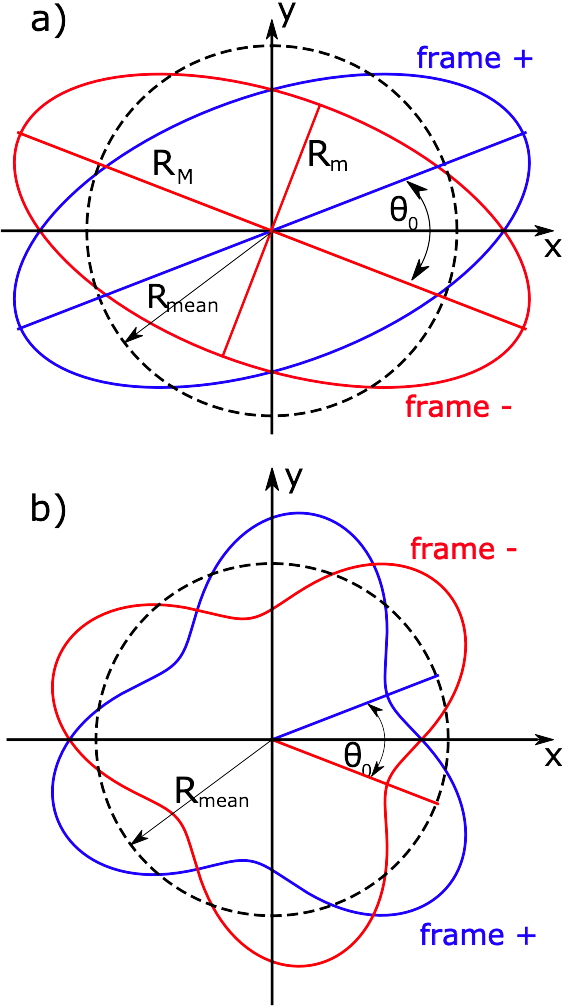}
    \caption{Upper (blue) and lower (red) frames, in the cases of the ellipse (a) and the clover (b), with the different associated geometric parameters $R_{\rm mean}$, $R_{\rm M}$, $R_{\rm m}$ and $\theta_0$.}
\label{fig:parameter}
\end{figure}

In this article, we extend previous studies by investigating the shape, the stability and the mechanical properties of a special group of minimal surfaces, spanning two identical {\bf non-axisymmetric} closed frames, which are contained in two parallel planes (the case of two different frames is described in Appendix \ref{appendixB}). The planar boundaries ${\cal C}_\pm$ ($+$ and $-$ denoting the upper and lower frame, respectively) are centered on the $z$ axis and given in polar coordinates $r_{\mathcal{C}\pm}\left(\theta\right)$. They are separated by a distance $h$ and rotated by an angle $\pm\theta_0/2$ around the $Oz$ axis. More specifically, elliptic and clover frames are used, as shown in Fig. \ref{fig:parameter}. 

Non-axisymmetric boundary conditions introduce an additional degree of freedom, the angle $\theta_0$ between the upper and lower frame. This brings very interesting new properties to the minimal surface. After quantify the influence of the angle $\theta_0$ on the existence of the minimal surface and the associated critical height, we pay special attention to their mechanical properties arising from a constant surface tension $\gamma$. We study the forces transmitted by the minimal surface to the frames. In particular, we show that besides the normal force, which pulls each frame towards the other one, non-axisymmetric shapes are also characterised by a measurable torsion torque, which tends to rotate the two frames back to the position $\theta_0=0$.

To analyse the different surface properties, we combine experiments with soap films, numerical simulations using the open source software Surface Evolver \cite{SE}, and theoretical modelling. The latter is a perturbation theory based on the catenoid solution. For all three approaches we study the shapes and the critical height of the minimal surfaces together with the resulting normal force and the torque on the frames. Very recently, Alimov \textit{et al.} \cite{alimov2020minimal,alimov2021soap} used an analogy between fluid dynamics of potential flow  and minimal surfaces to provide an iterative algorithm allowing to calculate the exact surface shape between two identical convex frames. However, their theory is restricted to systems without rotation (i.e. for $\theta_0=0$). In the following, it will be used as benchmark for our investigations together with the analytical predictions of the catenoid.    

\section{Materials and methods}

\subsection{Boundary conditions}

The shape of the minimal surface is represented in cylindrical coordinates with the vector position $\mathbf{r}$ given by $\left(x=r\left(\theta,z\right)\cos{\theta}, y=r\left(\theta,z\right)\sin{\theta},z=z\right)$.

At the top and bottom frames (i.e. for $z=h/2$ and $-h/2$ respectively), the boundary conditions make the surface have the same contour, up to a rotation of an angle $\theta_0$, a condition which can be written \footnote{The case of two different frames is described in Appendix \ref{different-frames}} $r_{\mathcal{C}\pm}(\theta)=r_{\mathcal{C}}(\theta\mp \theta_0/2)$
where $r_{\mathcal{C}}(\theta)$ is the unrotated contour defined via its Fourier decomposition
\begin{equation}
\label{eq:contour_frame}
r_{\mathcal{C}}\left(\theta\right)=R_{\rm mean}\left(1 + \sum_{k=2}^{+\infty} a_k\cos{\left(k\theta\right)} + b_k \sin{\left(k\theta\right)}\right).
\end{equation}
Note that the Fourier series starts at $k=2$ to ensure that the frame is centered, i.e. that $\langle x\rangle=\langle r\cos\theta\rangle=0$ and similarly with $y$.

As shown in Fig \ref{fig:parameter}, we use two different types of non-axisymmetric frames: elliptic shapes labelled $r_{\mathcal{C}} = r_{\rm e}$ and clover-type shapes labelled  $r_{\mathcal{C}} = r_{\rm cl}$.  

The elliptic frames are defined with an eccentricity $e$, major axis $2R_{\rm M}$, and minor axis $2R_{\rm m}$, with $e^2=1-(R_{\rm m}/R_{\rm M})^2$ and $R_{\rm M}=R_{\rm m}/\sqrt{1-e^2}$ (see Fig. \ref{fig:parameter}(a)). The polar equation of the non-rotated elliptic frame is given by
\begin{equation}
\label{eq:bound_el}
r_{\rm e}\left(\theta\right)=\frac{R_{\rm m}}{\sqrt{1-e^2\cos^2{\theta}}}.
\end{equation}
The upper and lower frames are rotated by $\pm \theta_0/2$, respectively. The first 2 coefficients from the Fourier-like expansion of the elliptic frame $r_{\rm e}$ are equal to
\begin{eqnarray}
&&R_{\rm mean}\left(e\right) = \frac 2\pi K[e^2]R_{\rm m},\\
&&a_2\left(e\right) = \frac{2}{e^2}\left(2 - e^2 - \frac{E[e^2]}{K[e^2]}\right),
\end{eqnarray}
where $K\left[x\right], E\left[x\right]$ are the complete elliptic integrals of the first and second kind (see Appendix \ref{PT-ellipses} for next order coefficients). Coefficients $a_k$ with odd indices and all $b_k$ are zero due to the symmetry at $\theta=\pi$ and $2\pi$ of ellipses.
Moreover, the perimeter of the ellipse is defined as $P=4E[e^2]R_{\rm M}$.

The clover frames are defined with  $a_3=-\epsilon$ and all other Fourier coefficients from Eq. (\ref{eq:contour_frame})  equal to zero. The polar equation of the clover frames is given by
\begin{equation}
\label{eq:bound_cl}
r_{\rm cl}\left(\theta\right)=R_{\rm mean}\left(1-\epsilon\cos{\left(3\theta\right)}\right).
\end{equation}

\subsection{Experiment}
\label{sec:Experiment}

\begin{figure}
    \centering
    \includegraphics[width=0.75\textwidth]{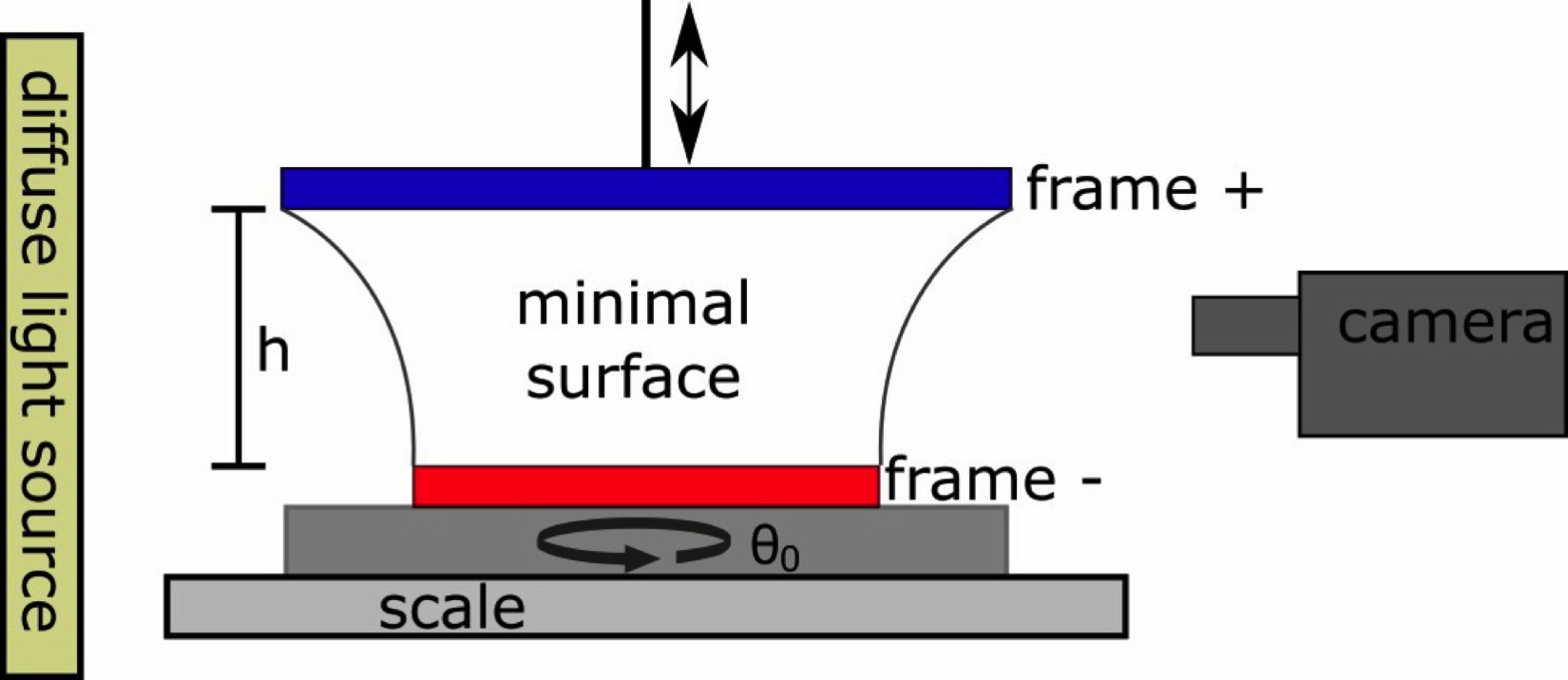}
    \caption{Scheme of the experimental set-up visualising soap films between 3D-printed identical frames whose separation is set by a vertical translation stage. The scale is used to measure the normal force $F_z$ and a goniometer imposes a rotation angle $\theta_0$ between the frames (the torque measuring device is not shown on this scheme).}
    \label{fig:setup_experiment}
\end{figure}

Minimal surfaces are studied using soap films held by 3D-printed frames with a set-up schematised in Fig. \ref{fig:setup_experiment}. The position of the lower frame remains fixed during the experiment, while the upper frame, attached to a vertical translation stage, can move at variable speed between controlled positions via a home written Labview program. The lower frame is fixed on a laboratory scale to measure the normal force $F_z$ acting between the frames. The ensemble is visualised from the side in front of a diffuse light source using a computer-controlled CCD camera with a spatial resolution of $50$ ${\rm \mu}$m. We performed experiments for two different elliptic frames: $e=0.866$ ($R_{\rm M}/R_{\rm m}=2$) and $e=0.97$ ($R_{\rm M}/R_{\rm m}=4$).  More details on the experimental set-up and protocols are given in Appendix \ref{AppendixA01}.

\subsection{Surface Evolver simulations}

Surface Evolver is an open source Finite Element program which represents a surface via vertices, edges and facets \cite{evolver}. Surface Evolver minimises the total energy by moving vertices of a defined shape in the oposite direction of the energy gradient. 
The total energy is in our case proportional to the total area $A$, defined here as the sum of the facets areas, since only a constant surface tension is considered. Vertices on the frame stay fixed at the position defined by the Eqs. (\ref{eq:bound_el}) or (\ref{eq:bound_cl}).

Our simulation procedure is similar to the procedure in the experiment. The height $h$ between the two frames is increased in small steps $\Delta h$ until the surface becomes unstable. To obtain the best precision in the critical height $h^\star$ we studied the eigenvalues $\lambda$ of the Hessian matrix $ \underline{ \underline{H}}$ close to the critical height, as described in detail in Appendix \ref{AppendixA02}. We investigate numerically elliptic frames of various eccentricities as well as clover frames.

The normal force $F_z$ and the torque $\Gamma_z$ along the axis passing by the two frame centers are related via

\begin{equation}
\label{eq:force_torque}
\gamma\dd A=\Gamma_z(\theta_0,h) \dd\theta_0 + F_z(\theta_0,h)\dd h.
\end{equation}

The force and the torque applied on the frame are the same along the surface. The infinitesimal increments $\dd A$, $\dd\theta_0$ and $\dd h$ are approximated by finite differences of two simulated surfaces with a small change in $h$ or $\theta_0$. The precision of these calculations depends strongly on the precision of the total area minimum, which, in turn, depends on the number of facets and the iteration process. 

\subsection{Implementation of the Alimov {\it et al.} method \cite{alimov2021soap}}

We implemented the algorithm described in \cite{alimov2021soap} using Python code. The code correctness of the implementation was tested by comparing the critical heights, critical areas and the shape parameters to the values given by Alimov {\it et al.} \cite{alimov2021soap} in their Supplemental Materials. All values of the table were reproduced with a relative difference smaller that $5.10^{-5}$. The authors introduce a parameter called $\Psi_B$ that we found directly proportional to the vertical component of the force acting on each frame.  The convergence criterion used in the paper, which is based on the stability of the second coefficient $C_2$ in the Laurent series expansion had to be generalised for the clover case, for which $C_2$ remains zero.

\section{Perturbation theory}
Here we present a perturbation approach to approximate minimal surfaces close to the catenoid. The two frames can therefore be of different shapes, in contrast to the  model presented by Alimov \textit{et al.} \cite{alimov2021soap}.
\\The computation of the perturbation theory concerning the non-axisymmetric minimal surface consists in solving the minimal surface differential equation, corresponding to a vanishing mean curvature. We start by recalling the form of this partial differential equation in cylindrical coordinates, then we present in the general case the perturbative scheme we have developed to solve this equation and we discuss the conditions of the existence of a surface in the case of identical boundary conditions (the case of two different frames is described in Appendix D.1). Finally we show how the perturbative approach allows us to calculate the normal force and the torque exerted by the minimal surface on the contours.

\subsection{Notations} \label{PT-Ge}

In what follows we use the common notations $r_i\left(\theta, z\right)=\partial r/\partial i$ and $r_{ij}\left(\theta, z\right)=\partial^2 r/\partial i\partial j$ with $i,j=\theta,z$. We can derive the following partial differential equation resulting from the vanishing mean curvature problem (see Appendix \ref{appendixA})
\begin{eqnarray}
\nonumber
r_\theta^2+r\left[r\left(1+r_z^2\right) - r_{\theta\theta}\left(1+r_z^2\right) - r^2 r_{zz} \right.\\
\left.+ r_\theta\left(2 r_z r_{z\theta}-r_\theta r_{zz}\right)\right]=0.
\label{eq:equadiffgene}
\end{eqnarray}

If we suppose that the minimal surface is invariant by rotation (corresponding to $r_{\theta}=r_{\theta\theta}=0$), we obtain $1+r_z^2 - r r_{zz} = 0$. The solution of this equation is simply the symmetric catenoid for $z \in [-h/2 ; h/2]$, which was given in Eq. (\ref{eq:catenoid_shape}).

It is worthwhile to mention that the Eq. (\ref{eq:equadiffgene}) corresponds to the Euler-Lagrange equations minimising the surface formula
\begin{equation}\label{S}
S[r(\theta,z)]=\int_{-h/2}^{h/2}{\rm d}z\oint {\rm d}\theta\sqrt{r^2+r_\theta^2+r^2r_z^2}.
\end{equation}
This Lagrangian interpretation of this optimisation problem allows to anticipate the conservation of a pseudo energy function, namely the fact that
\begin{equation}\label{conservationofa}
    a_h=\oint\frac{{\rm d}\theta}{2\pi}\frac{r^2+r_\theta^2}{\sqrt{r^2+r_\theta^2+r^2r_z^2}}
\end{equation}
is a constant independent of $z$ (along the true {\it minimal} surface). If the solution for the axisymmetric catenoid is plugged in this equation, one recovers for $a_h$ the neck radius $a_c$ introduced in Eq. (\ref{eq:catenoid_shape}).

\subsection{General perturbation solution}

Here we solve Eq. (\ref{eq:equadiffgene}) considering boundary contours which are small perturbations of a circle. We consider the Ansatz
\begin{equation}
r\left(\theta, z\right)=a\cosh{\left(\frac{z}{a}\right)}\left(1 + f\left(\theta, z \right)\right),
\label{eq:r}
\end{equation}
with $f(\theta, z)$ a perturbative term. In this case $R_0(z)=a\cosh{\left(z/a\right)}$ is a reference catenoid (with $a$ to be determined). Rewriting  Eq. (\ref{eq:equadiffgene}) using $f$ and restricting it to the first order in $f$, we end up with a linear differential equation of the form \footnote{This equation can also be obtained from the Euler-Lagrange equation of the second-order expansion of Eq. \ref{S} with respect to $f$, which reads : $S[f]=\frac{a}{2}\int_{-h/2}^{h/2}\dd z\oint \dd\theta[f_\theta^2+(af_z)^2-2f^2\text{sech}^2(z/a)]$ plus boundary terms.}
\begin{equation}
\label{equadifordre1}
2 f\left(\theta, z\right) + \cosh^2{\left(\frac{z}{a}\right)} \left(f_{\theta\theta}\left(\theta, z\right) + a^2 f_{zz}\left(\theta, z\right)\right)=0.
\end{equation}
To solve (\ref{equadifordre1}), let us decompose $f$ into a Fourier series 
\begin{equation}
\label{fouriergene}
f\left(\theta, z\right) = \alpha_0(z)+\sum_{k=1}^{+\infty}[\alpha_k\left(z\right)\cos{\left(k\theta\right)}+\beta_k\left(z\right)\sin{\left(k\theta\right)}].
\end{equation}
Introducing this expansion in Eq. (\ref{equadifordre1}), 
the equation separates each mode, and the equation for mode $k$ can be written as
\begin{eqnarray}
\nonumber
\label{equadifalpha}
\left(2 - k^2\cosh^2{\left(\frac{z}{a}\right)}\right)\alpha_k\left(z\right) + a^2\cosh^2{\left(\frac{z}{a}\right)} \alpha_k^{\prime\prime}\left(z\right)=0,\\
\end{eqnarray}
(and the same equations for the $\beta_k$-s), 
which can be solved independently for each order $k$, with different boundary conditions for $\alpha_k$ and $\beta_k$.

It is now possible to introduce the boundary conditions for the minimal surface
\begin{eqnarray}
\label{bound}
r\left(\theta,\pm h/2\right)=r_{\rm \mathcal{C}\pm}\left(\theta\right),
\end{eqnarray}
which, in the case of two identical frames, leads to
\begin{eqnarray}
\label{boundary0}
&&a\cosh{\left(\frac{h}{2a}\right)}\left(1+\alpha_0(\pm\frac{h}{2})\right) = R_{\rm mean},\\
\label{boundaryalpha}
&&\alpha_k\left(\pm h/2\right) = a_k\cos{\left(k\frac{\theta_0}{2}\right)}\mp b_k\sin{\left(k\frac{\theta_0}{2}\right)},\\
\label{boundarybeta}
&&\beta_k\left(\pm h/2\right) = \pm a_k\sin{\left(k\frac{\theta_0}{2}\right)}+b_k\cos{\left(k\frac{\theta_0}{2}\right)}.
\end{eqnarray}
The Ansatz (\ref{eq:r}) embodies a small departure (with $|f|\ll 1$) from a reference catenoid characterized by its parameter $a$. A convenient choice for this reference catenoid is obtained by taking $\alpha_0(\pm h/2)=0$ : In this case the boundary radius of this reference catenoid is given by the mean radius $R_{\rm mean}$ of the actual elliptic or clover contour (compare Eqs. (\ref{boundary0}) with (\ref{eq:neck_radius})). As the solution for $\alpha_0(z)$ is known and equal to $\alpha_{0}^{(1)}\tanh(z/a)+\alpha_{0}^{(2)}((z/a)\tanh(z/a) -1)$ (with $\alpha_{0}^{(1,2)}$ some constants), the constraints $\alpha_0(\pm h/2)=0$ imply $\alpha_0(z)=0$. It is worth noting that this approximation entails therefore that the $\theta$-averaged radius $(2\pi)^{-1}\oint r{\rm d}\theta$ describes the reference catenoid.

For a given frame we therefore need to calculate the coefficients  $a_k$ and $b_k$ up to an order of Fourier expansion $k_M$, and then solve the differential Eq. (\ref{equadifalpha}) 
with the boundary conditions (\ref{boundaryalpha}) and (\ref{boundarybeta}) (see Appendix \ref{appendixB} for more details).  
Since the shape is completely defined with $a_k$ and $b_k$, all properties of the surface, like normal forces and torque can be calculated now.  
Eq. (\ref{boundary0}) (with $\alpha_0=0$)  generalizes the condition of existence of the catenoid to more general frames leading to the theoretical critical height $h_{\rm theo}^\star$
\begin{equation}
h_{\rm theo}^\star = \mu^\star R_{\rm mean}.
\end{equation}
When $h < h_{\rm theo}^\star$, Eq. (\ref{boundary0}) has two solutions for $a$, just like the catenoid case. By comparing Eqs. (\ref{boundary0}) and (\ref{eq:catenoid_shape}), one observes an equivalence between $R_{\rm mean}$ and $r_c$ of the catenoid Eq. (\ref{eq:catenoid_shape}). As the zero order Fourier coefficient $R_{\rm mean}$ is independent of $\theta_0$ or $e$, at this level of perturbation, the criterion of existence is independent of the angle $\theta_0$ and $e$.

The case of a minimal surface supported by two circles of different radius being well known (asymmetric catenoid \cite{Salkin}), the perturbative theory can easily be extended to the case of a minimal surface supported by two different frames (see Appendix \ref{appendixB}).

\subsection{Force and torque predictions}
\label{force:theo}

The force and the torque applied by the minimal surface on the frames can also be computed thanks to the perturbation theory. The elementary surface tension force acting on a length element $\dd{\bf \ell}{\bf t}_{\cal C_\pm}$ along a contour ${\cal C_\pm}$ is given by $\dd{\bf F}_{\mathcal{C}_\pm}=2\gamma{\bf t}_{\mathcal{C}_\pm}\wedge{\bf n}\dd{\bf \ell}$ where ${\bf n}$ is the vector normal to the surface and ${\bf t}_{\cal C_\pm}$ is the tangent to the frames ${\cal C_\pm}$. 

It is possible to give a geometrical interpretation of this force. At equilibrium, the force can be computed on any closed contour ${\cal C}$ not reducible to a point by calculating $2\gamma \oint_{\cal C} {\bf n}\wedge{\bf t}_{\cal C}\dd{\bf \ell}$, where ${\bf n}$ is the normal to the surface, ${\bf t}_{\cal C}$ is the tangent to the contour and $\dd\ell$ is an element of length tangent to the contour. There is a particular contour, ${\cal C}_{\rm neck}$, for which at any point the normal to the surface is horizontal, which generalises the notion of the "neck" for the catenoid. In the case of a catenoid it is a circle in the mid-plane, whereas for arbitrary frames this contour is more complex, generally not restricted to the mid-plane and in some cases, can be discontinuous. When this contour is fully contained in the minimal surface, the $z$ component of the force can be calculated using the simple expression (see also Appendix \ref{appendixC})
\begin{equation}
\dd{F}_z=2\gamma P_{{\cal C}_{\parallel}},
\label{eq:projected_length}
\end{equation}
where $P_{{\cal C}_{\parallel}}$ is the perimeter of ${\cal C}_{\parallel}$, the projection of ${\cal C}_{\rm neck}$ onto the median plane. $F_z$ is therefore a direct measure of the perimeter of the projection onto the median plane of the locus of points where the normal to the surface is horizontal.

The torque can be calculated using $\dd{\Gamma}_{\mathcal{C}_\pm}={\bf r}_{\mathcal{C}_\pm}\wedge{\bf\dd{ F}}_{\mathcal{C}_\pm}$, leading for the $z$ component to
\begin{eqnarray}
\Gamma_{\mathcal{C}_\pm,z} &=& 2\gamma\int_{\mathcal{C}_\pm} r_{\mathcal{C}_\pm}\left(\theta\right)\left(\cos{\theta}\dd {F}_{{\mathcal{C}_\pm},x} - \sin{\theta}\dd{F} _{{\mathcal{C}_\pm},y}\right).
\end{eqnarray}

\section{Results and discussions}

In the following, we systematically compare the measured and calculated quantities of the soap film experiments (index "exp") and of the Surface Evolver simulations (index "SE") with the predictions of the perturbation theory (index "P") and of the algorithm provided by Alimov {\it et al.} \cite{alimov2021soap} (index "A"). We analyse first the stability of the continuous minimal surface  (Section \ref{sec:Resultsh*}) and their shapes (Section \ref{sec:ResultsShape}), investigating in detail, in the cases of elliptic and clover frames, the effect of the angle $\theta_0$ between the frames on the critical height $h^\star$. We then turn to the analysis of the normal force (Section \ref{sec:ResultsNormForce}) and the torque (Section \ref{sec:ResultsTorque}) exerted on the frames by the film.

\subsection{Critical height}
\label{sec:Resultsh*}

\begin{figure*}
\includegraphics[width=0.9\textwidth]{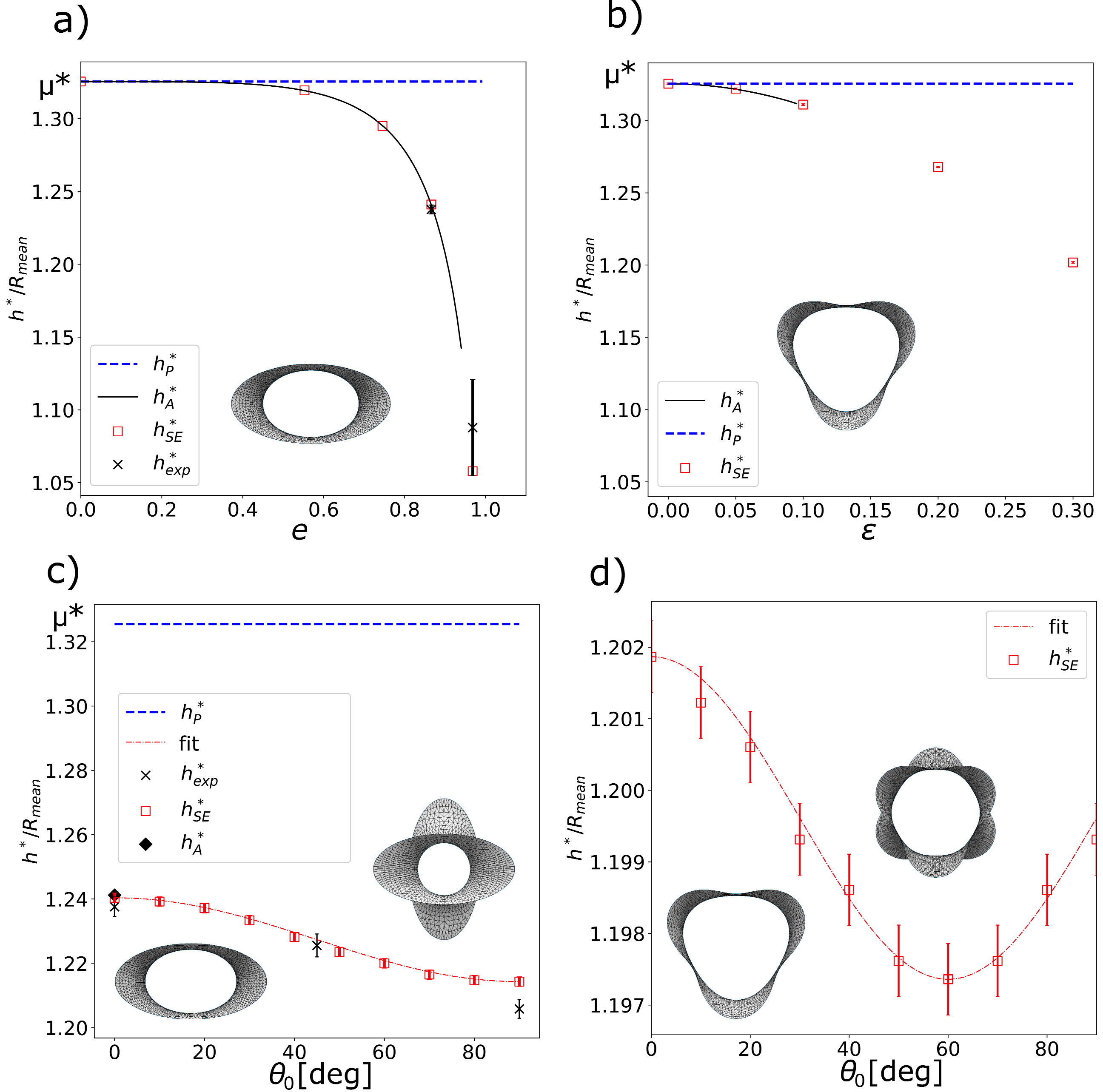}
\caption{Normalized critical height ($h^\star/R_{\rm mean}(e)$) of minimal surfaces with elliptic a) and c) and clover frames b) and d) obtained with Surface Evolver simulations $h^\star_{\rm SE}$ (\textcolor{red}{$\square$}), experiments $h^\star_{\rm exp}$ ($\times$), the perturbation theory $h^\star_{\rm P}$ (dashed line) and the the Alimov algorithm $h^\star_{\rm A}$ (solid line). a) and b) show how $h^\star$ changes with $e$ or $\epsilon$ for $\theta_0=0^\circ$. b) and c) show the dependence on $\theta_0$ for constant values of $e=0.866$ in c) and $\epsilon=0.3$ in d). The red dashed line in figure c) and d) is an empirical fit with a cosine function.}
\label{fig:critical_height_all}
\end{figure*}

We first focus on minimal surfaces with elliptic boundary conditions. Figure \ref{fig:critical_height_all}a displays the critical height $h^\star$ normalized by the average radius $R_{\rm mean}(e)$ as a function of eccentricity $e$ for $\theta_0=0$. Experimental measurements, SE simulations and exact theory (Alimov {\it et al.} method \cite{alimov2021soap}) are in perfect  agreement for all $e$. This confirms the validity of our experimental and numerical protocols. The perturbation theory also agrees very well with the Alimov {\it et al.} method \cite{alimov2021soap} over a wide range of $e$.

\begin{figure}[h!]
    \centering
    \includegraphics[width=0.49\textwidth]{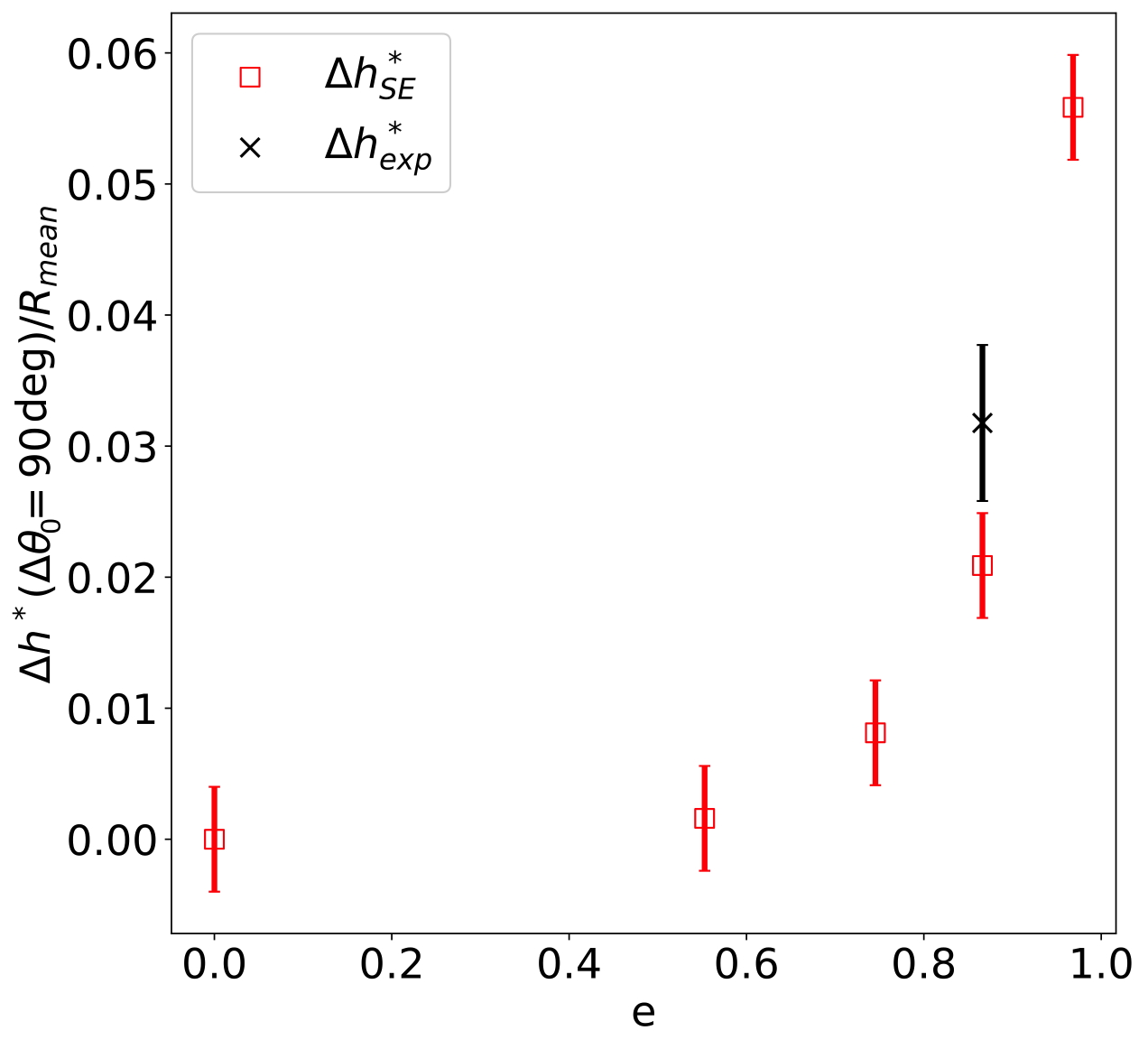}
    \caption{Normalized maximal variation of $h^\star$, $\Delta h^\star/R_{\rm mean}$ for different eccentricities $e$. Only simulations and experiments, which are able to show these changes, are presented.}
    \label{fig:dhc_ellipse_theta}
\end{figure}

The next step is to evaluate the impact of the angle $\theta_0$ between the two elliptic frames (see Fig.\ref{fig:parameter}(a) on $h^\star$ with the different approaches. As shown in the Fig. \ref{fig:critical_height_all}(c), experiments and SE simulations show a very small variation of the critical height with the angle $\theta_0$ (less than 2\% for $e=0.866$), compatible with a $\cos{\left(2\theta_0\right)}$ variation. Experimentally, this small difference of $h^\star$ for different $\theta_0$ is only evidenced by computation of the statistical average over repeated measurements. This variation is not predicted by the perturbative theory, as shown by equation \ref{boundary0}, which systematically overestimates $h^\star$ (see Fig. \ref{fig:critical_height_all} c)). Figure \ref{fig:dhc_ellipse_theta} displays the maximum normalised variation of $h^\star$ between the different angles $\theta_0$ expressed {\it via} $\Delta h^\star=\left(h^\star\left(\theta_0=0\right)-h^\star\left(\theta_0=90\right)\right)/R_{\rm mean}$ as a function of eccentricity $e$. The impact of the angle $\theta_0$ on $h^\star$ increases quite naturally with the increase of $e$.

We have extended numerically this study to other boundary conditions. Figure \ref{fig:critical_height_all}(b) and (d) display the results obtained for clover frames. Again, a very small but significant variation of $h^\star$ is observed with $\theta_0$. Relative changes are even smaller than for the ellipses, $0.4\%$, for almost the same critical height for $\theta_0=0$ ($e=0.866$ and $\epsilon=0.3$). As expected from the shape of the contours, we observe a three fold symmetry in the case of the clover in good agreement with a cosine variation (see Fig. \ref{fig:critical_height_all}(d)). The comparison with theoretical models is similar to the case of ellipses. Again $h^\star_{\rm P}$ is constant for all $\epsilon$, Fig. \ref{fig:critical_height_all}(b). The Alimov {\it et al.} method \cite{alimov2021soap}, which can be applied only to convex shapes (i.e. for $\epsilon < 0.1$ in our case), predicts again very well the variation of $h^\star$ with $\epsilon$ (see Fig. \ref{fig:critical_height_all}). For small $\epsilon$ values, all approaches are in good agreement with the SE simulations. 

In conclusion, concerning the critical height $h^\star$, we have demonstrated both experimentally and numerically, that a small but significant dependence of $h^\star$ on the angle $\theta_0$ between the frames exists for different contour shapes (ellipses and clovers). This variation is not predicted by the perturbative theory, and, to our knowledge, no alternative theoretical prediction exists at this stage which captures this observation. Interestingly, we can see that the perturbative theory predicts a constant critical height independently of the number of Fourier coefficients taken into account. The variation of $h^\star$ is therefore intrinsically linked to the non-linear character of the partial differential equation (\ref{eq:r}).

\subsection{Shape description}
\label{sec:ResultsShape}

\begin{figure}
        \centering
        \includegraphics[width=0.75\textwidth]{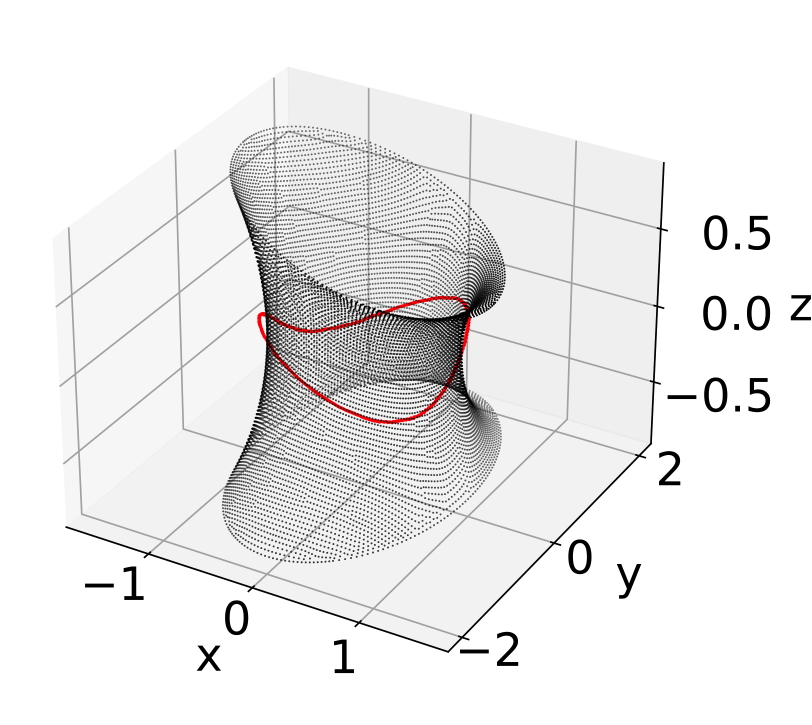}  
        \caption{Shape of a minimal surface held by elliptic frames for one example with $\theta_0=45^\circ$, $h/R_{\rm mean}=1.02$ and $e=0.866$,  obtained by SE simulations (all vertices are shown). The red solid line is the neck contour of this surface.}
        \label{fig:shape_3D_surface_evolver}
\end{figure}

In this section we focus on the shape of the minimal surfaces for $h<h^\star$. Figure \ref{fig:shape_3D_surface_evolver} shows the example of a minimal surface obtained via SE simulations for  elliptic frames at an angle $\theta_0=45^\circ$, $h/R_{\rm mean}=1.02$ and $e=0.866$. The surface is represented by plotting the vertices of the finite element mesh. 

\begin{figure*}
 \centering
 \includegraphics[width=0.9\textwidth]{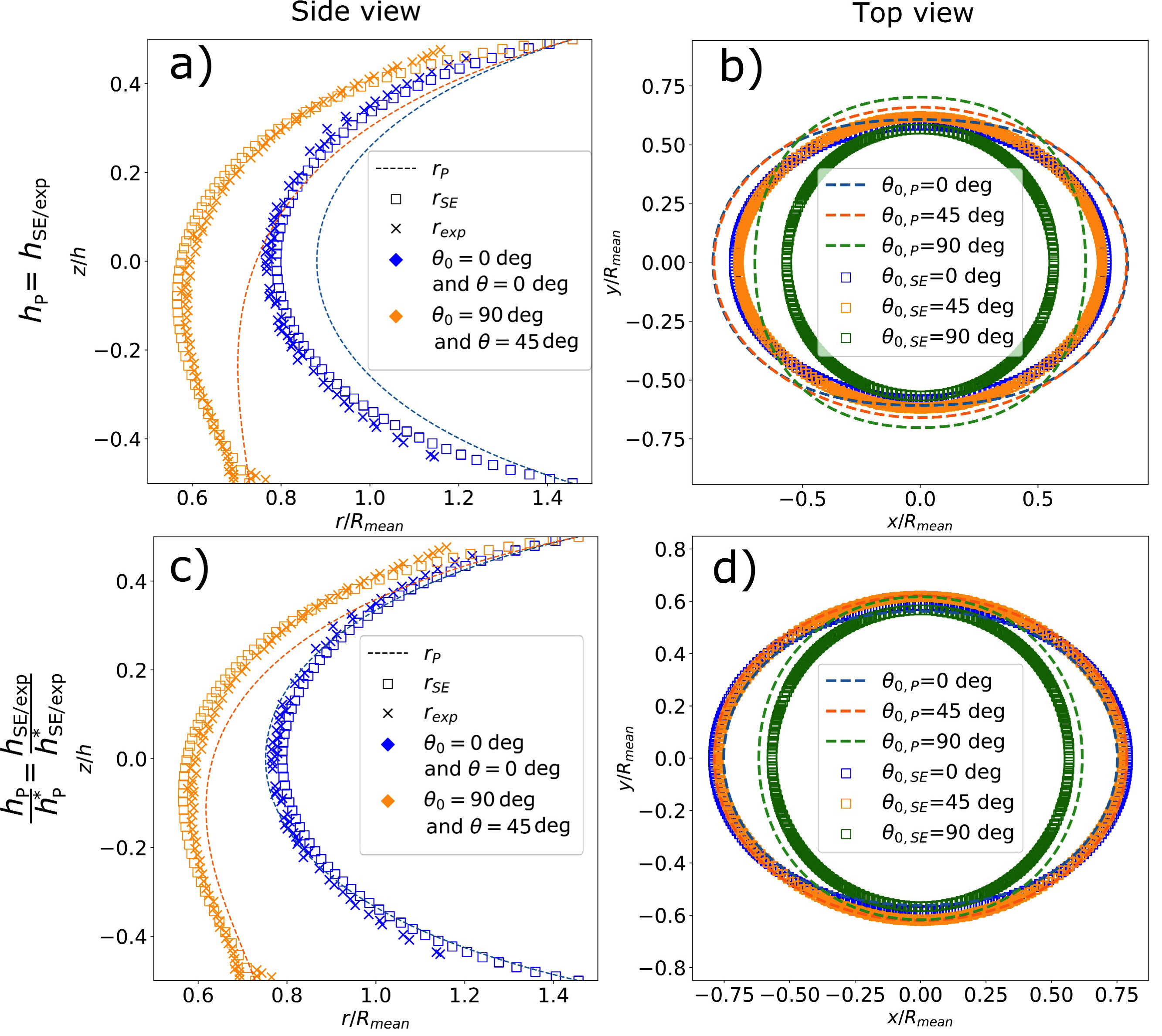}
 \caption{(Left) Side view (vertical profile in the symmetry plane of the surface defined by $\theta$) and (Right) top view (projected neck contour) of a minimal surface spanning elliptic frames with $e=0.866$ obtained with SE simulations $r_{\rm SE}(z)$, the perturbation theory $r_{\rm P}(z)$ and the soap film experiment $r_{\rm exp}(z)$ for two different angles $\theta_0$ at a height of $h=1.2R_{\rm mean}$. On the right, the dashed lines are the predicted neck contours of the perturbation theory and the open squares are the vertices closest to the neck contour of the Surface Evolver simulation.}
 \label{fig:shapes_all}
\end{figure*}

In order to make a quantitative comparison of our results, we extracted the experimental, numerical and theoretical profiles corresponding to the intersection of the surface with a vertical plane passing through the axis of rotation, and making an angle $\theta$ with the x-axis (see Fig. \ref{fig:shapes_all} for 2 values of $\theta_0$ and $\theta$). Experiments and simulations are in very nice agreement for both profiles. Far from the critical point ($h\ll h^\star$), the perturbative theory is in very good agreement with the experimental and numerical results (data not shown here, see Appendix \ref{appendixB}). Near the critical point ($h\leq h^\star$), the agreement is less convincing (See Fig. \ref{fig:shapes_all}(a) and (b), and the perturbative theory clearly overestimates the radius $r(\theta,z)$. As we have shown before, the perturbative theory does not predict a variation of the critical height $h^{\star}$ with the angle $\theta_0$. This means that this variation appears in the non-linear terms of the differential equation (\ref{eq:equadiffgene}), whereas there are of course corrections to the profile that depend on $\theta_0$ at the linear order. Rather than comparing the experimental, numerical and theoretical profiles for the same value of $h/R_{\rm mean}$, as done before, we therefore compared them for the same value of $h/h^\star$, i.e. at the same relative distance to the critical point. The results presented in Fig. \ref{fig:shapes_all}(c) and (d) show a good agreement for both types of profiles (top and side views), which demonstrates that the perturbative theory describes quantitatively very well the shape of the surface, even very close to the critical point, provided that variations of $h^\star$ are taken into account. 

All profiles show a point where $r(z)$ is minimal and the normal to the surface is horizontal corresponding to the Top view (see Fig. \ref{fig:shapes_all}). The location of these points generalises the notion of the neck contour introduced in the case of the catenoid. For non-axisymmetric frames, it is a closed loop, which is not necessarily planar. An example is represented by the red solid line in Fig. \ref{fig:shape_3D_surface_evolver}. For large $\theta_0$ and small $h$, the neck contour may lie partially outside the surface. Mathematically, the neck contour is the curve that minimises its projected perimeter in the $Oxy$ plane ${\cal C}_{\parallel}$. From a physical point of view, the neck contour has a particularly interesting property. The total force exerted by the minimal surface on the frame gives directly access to the projected length of the neck contour in the $Oxy$ plane ${\cal C}_{\parallel}$. 

\subsection{Normal forces on frames}
\label{sec:ResultsNormForce}

\begin{figure}
    \centering
    \includegraphics[width=0.75\textwidth]{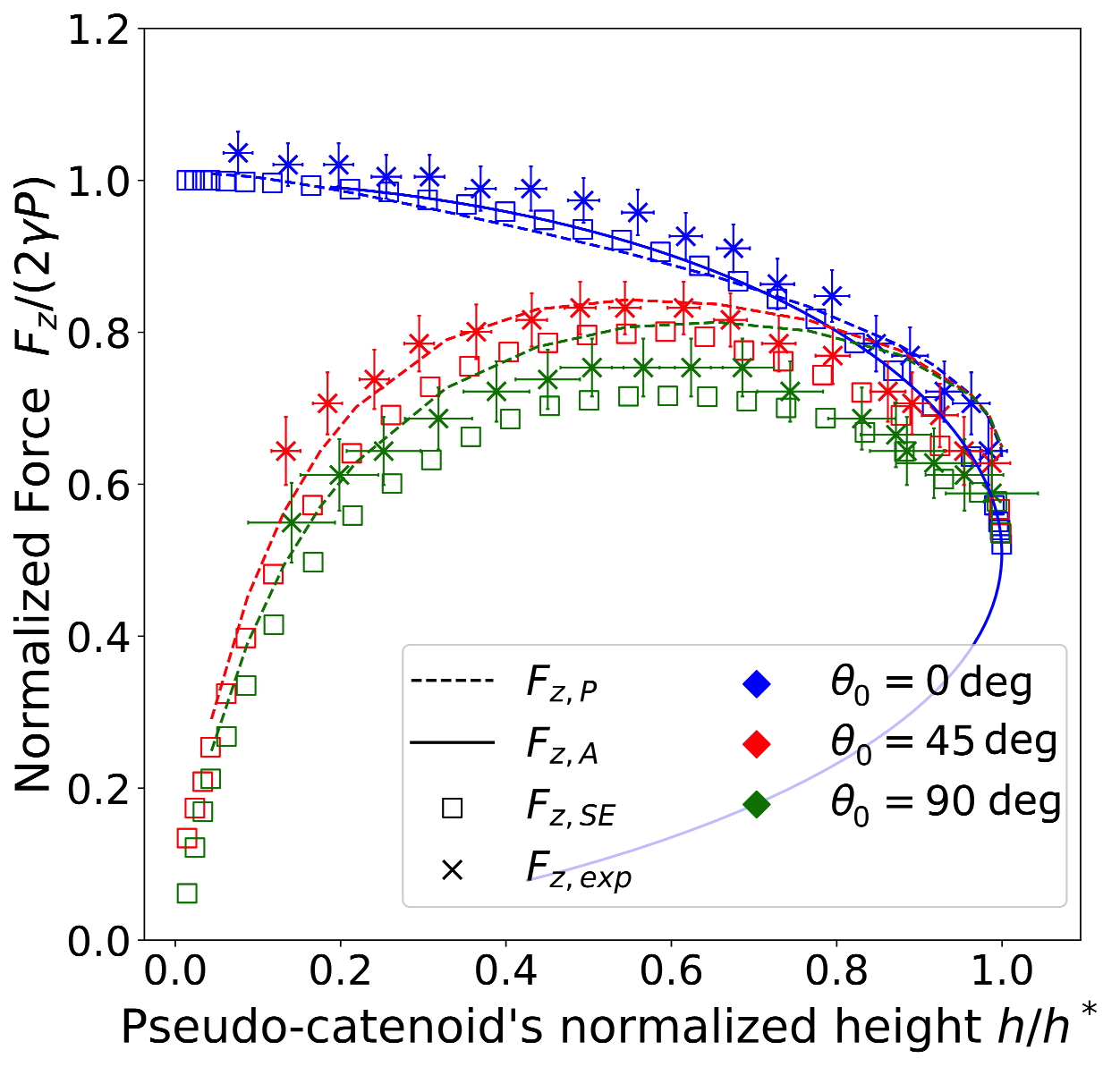}
    \caption{Normalized normal force $F_z/2\gamma P$ acting on elliptic frames for different heights $h$ and angles $\theta_0$, obtained with SE simulations $F_{z,{\rm SE}}$, soap film experiments $F_{z,{\rm exp}}$, the perturbation theory $F_{z,P}$ and the Alimov {\it et al.} method \cite{alimov2021soap} $F_{z,A}$.}
    \label{fig:forces}
\end{figure}

We now discuss the results for the normal force $F_z$ exerted by the soap film on the elliptic frame (Section \ref{sec:Experiment}). $F_z$ depends on the distance $h$ and the angle $\theta_0$ between the two frames, as shown in Fig. \ref{fig:forces}. In the Alimov {\it et al.} method, one has to calculate a value (called $\Psi_B$ in \cite{alimov2021soap}), which is proportional to the projected neck perimeter ${\cal C}_{\parallel}$ and thus to $F_z$. The Alimov algorithm thus allows to compute the force $F_{z,A}$ , which will serve as a reference case. For $\theta_0=0^\circ$,  $F_z$ is monotonically decreasing with $h$ until it reaches $h=h^\star$. For very small $h$ and $\theta_0=0^\circ$ the normal force is maximal   $F_{z,{\rm Max}}=2\gamma P$, where $P$ is the perimeter of the frame. 

Also for $h=0$, an infinitely small angle $\theta_0\neq 0$ is sufficient for the force to vanish, $F_z=0$. This discontinuity of $F_z\left(h=0,\theta_0\right)$ for $\theta_0=0$, which may seem surprising, simply reflects the fact that for $h=0$, as soon as there is a non-zero angle between the frames, the minimal surface is made up of horizontal films (see Figure \ref{fig:h=0} in Appendix \ref{appendixC}) and the force is null.

For $\theta_0\neq 0$, the force increases with $h$ until it reaches a maximum and decreases again up to $h=h^\star$. This non-monotonic behavior of the force is discussed in Appendix \ref{forcemax}. The experiments, the SE simulations and the two theories are in good agreement with each other for all the boundary conditions indicated, and in particular describe well the evolution of the position of this maximum with the angle $\theta_0$. This was expected, since ${\cal C}_{\parallel}$ in Fig. \ref{fig:shapes_all} shows a good agreement between the different methods.

\begin{figure}
    \centering
    \includegraphics[width=0.75\textwidth]{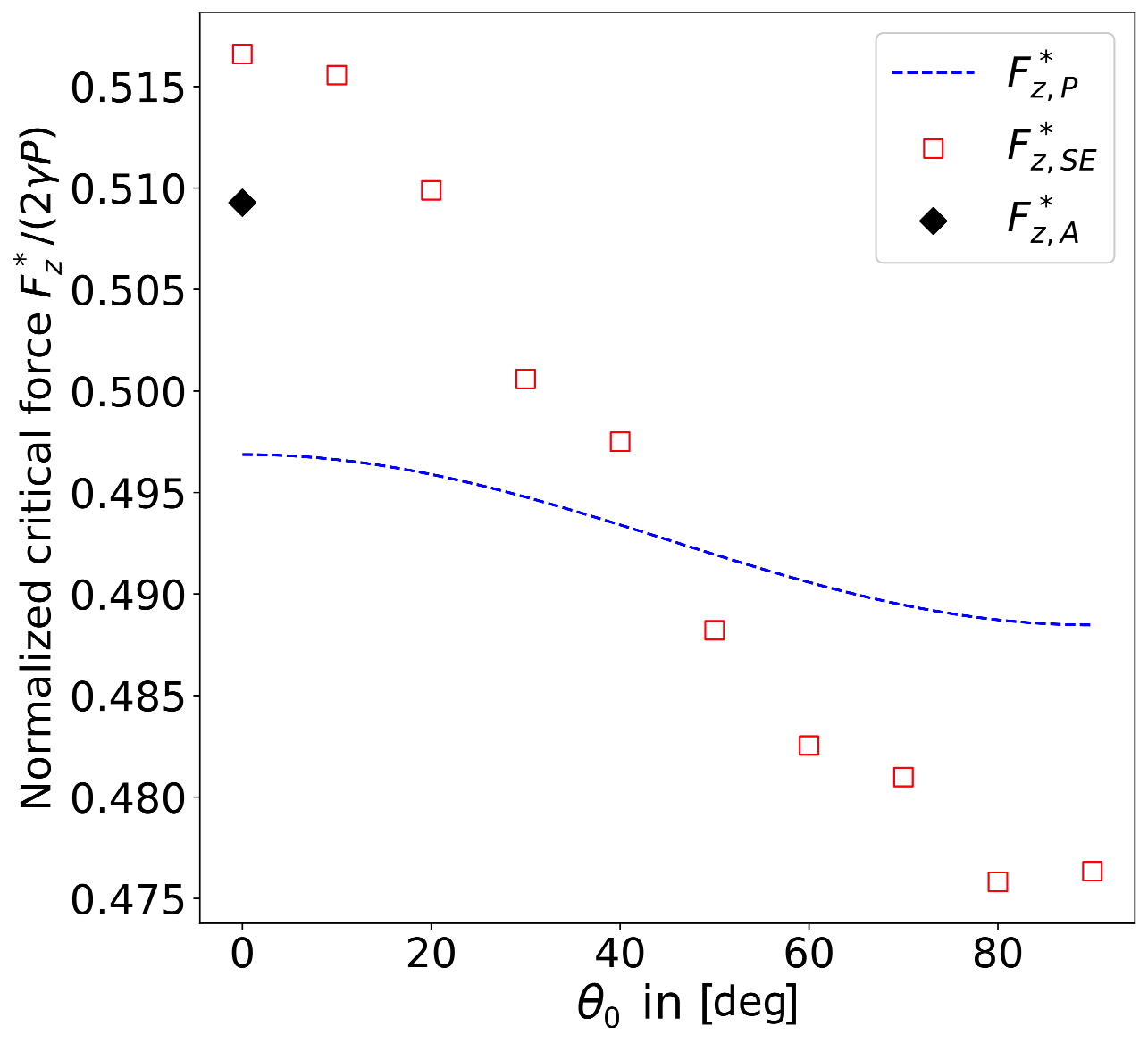}
    \caption{Variation of the critical, normalized normal force on the elliptic frames at $h=h^\star$ with $\theta_0$, obtained with SE simulations $F_{z,{\rm SE}}^\star$, the perturbation theory $F_{z,{\rm P}}^\star$ and the Alimov {\it et al.} method \cite{alimov2021soap} $F_{z,A}^\star$. We remind the reader that $h^\star$ is independent of $\theta_0$ in the perturbation theory. In the SE simulations all $F_{z,{\rm SE}}^\star$ belong to different $h^\star$.}
    \label{fig:force_critical_point}
\end{figure}

At the critical point, the results seem to suggest that the normal forces for different angles $\theta_0$ converge to the same value. In Fig. \ref{fig:force_critical_point} we therefore plot the normal force at the critical point $F_z^\star$ as a function of $\theta_0$. The difference between the forces is indeed small but clearly observable with the SE simulations. The perturbation theory also predicts a change in the force but with a much smaller variation with $\theta_0$.

In summary, sufficiently far from the critical point, experiments, SE simulations, perturbation theory and the Alimov {\it et al.} method \cite{alimov2021soap} show excellent agreement in the prediction of the normal force exerted on the elliptic frames by the minimal surface. At the critical point, excellent agreement in the dependence of the force on the angle $\theta_0$ is obtained only between the experiments and the SE simulations, while the perturbation theory predicts only the general trend.

\subsection{Torque on frames}
\label{sec:ResultsTorque}

 An important feature of minimal surfaces spanning non-axisymmetry frames is that, despite being fully liquid, they are able to exert an important torque $\Gamma_z$ on the frames. Here we investigate how this torque depends on the different frame geometries.

\begin{figure}[h!]
    \centering
    \includegraphics[width=0.75\textwidth]{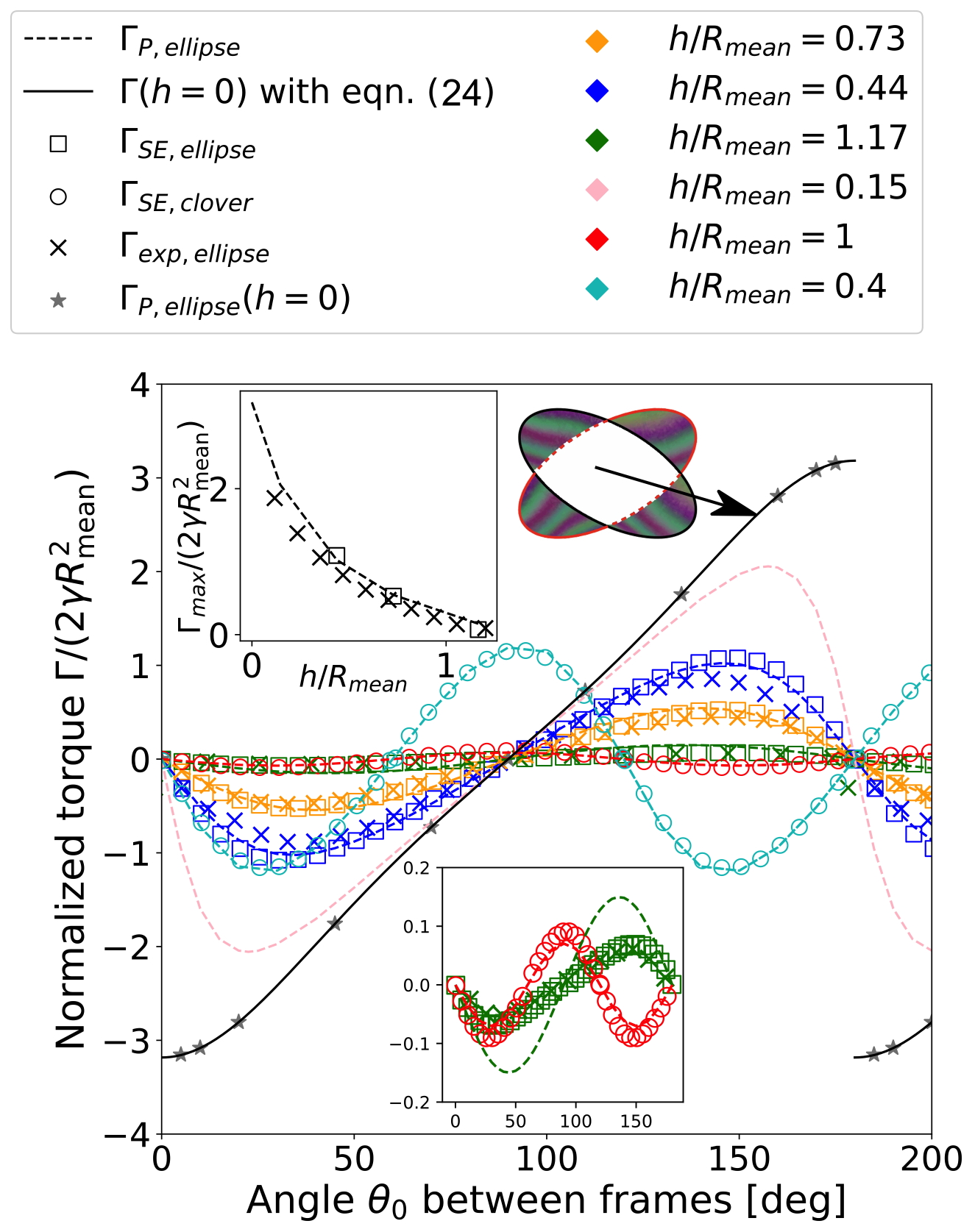}
    \caption{Torque on elliptic and clover frames for different $h$ (see color code in the legend) and $\theta_0$: experimental values $\Gamma_{\rm exp}$ for elliptic frames ($\times$), Surface Evolver simulations $\Gamma_{\rm SE}$ for elliptic frames ($\square$) and for clover frames ($\circ$), perturbative theory $\Gamma_{\rm P}$ (dashed lines) and an analytical solution for $h=0$ with Eq. (\ref{eq:torque_h=0}) (black solid line). The excentricity of the elliptic frame is $e=0.866$ and $\epsilon=0.3$ for the clover frame. The lower inset shows a zoom for small $\Gamma_z$ (same axis). The upper left  inset shows how the maxima of $\Gamma_z$ depend on $h$ for the case of ellipses. The upper right inset shows an example of the configuration of the horizontal  films for $h=0$.}
    \label{fig:torque_ellipse}
\end{figure}

Fig. \ref{fig:torque_ellipse} summarises all our experimental, numerical and theoretical results on the torque $\Gamma_z$ for elliptic and for clover frames at two different heights. The main figure plots the variation of the normalized torque $\Gamma_z/2\gamma R_{\rm mean}^2$ with the angle $\theta_0$. We first observe that the torque varies periodically, with a periodicity directly related to that of the frame: $180^\circ$ for elliptic frames and $120^\circ$ for clover. 

As expected, the torque vanishes when the axes of symmetry of the upper and lower frame are parallel, corresponding to a physically stable state ($\theta_0=0^\circ$ or $180^\circ$ for the ellipses and $\theta_0=0^\circ$ or $120^\circ$ for the clover) or a physically unstable state ($\theta_0=90^\circ$ for the ellipses and $\theta_0=60^\circ$ or $180^\circ$ for the clover). The torque thus presents two extrema, a maximum and a minimum over one period. The angle $\theta_0$ at which they appear depends on $h$, as it is clearly shown in Fig. \ref{fig:torque_ellipse}. For $h$, the torque has a linear behavior with a torsion constant which depends on $h$ and which increases with $h$ until diverging for $h=0$. Interestingly, the limit $h=0$ allows an analytical calculation of the torque since in this case the minimal surface is composed of pieces of plane films connected by points (see top right inset of Fig. \ref{fig:torque_ellipse}). In the case of elliptic frames, one obtains (see also Appendix \ref{appendixC})
\begin{equation}
\label{eq:torque_h=0}
    \Gamma_z(\theta_0, h=0)=8\frac{\left( R_{\rm m}e\right)^2\cos{(\theta_0)}}{(e^2-2)^2 - e^4\cos^2{(\theta_0)}}.
\end{equation}
This analytical result is represented by the black curve in Fig. \ref{fig:torque_ellipse} and is in perfect agreement with the perturbative theory (black $\star$). It shows that the torque presents a discontinuity at $\theta_0=0$ for $h=0$. The agreement between SE simulations and the perturbation theory is very good for low $h$, for both elliptic and clover frames. As one approaches the critical point ($h\leq h^\star$), the torque becomes very small and the agreement with the perturbation theory is less good. It can be shown that if we take into account the variation of $h^\star$ with $\theta_0$, by comparing the experimental and numerical values and the theory for the same values of $h/h^\star$ (as we did in Fig.  \ref{fig:shapes_all}) the agreement is much better. The top left inset of Fig. \ref{fig:torque_ellipse} plots how the maxima of the torque $\Gamma_{max}$ vary with $h$, showing again a very good quantitative agreement between the experiments, the SE simulations and the perturbation theory.

As shown in Fig. \ref{fig:torque_ellipse}, in the case of elliptic frames, the torque tends to align the two frames along the same axis. This shows that in this case the surface energy is minimal. We do not know if it is possible to generalize this property to the general case of two identical frames of arbitrary shape, even if it is obvious that it is correct when $h\rightarrow 0$.

In summary, far from the critical point, experiments, SE simulations and perturbation theory show excellent agreement in the description of the torque exerted by minimal surfaces on non-axisymmetric frames. Close to the critical point, the agreement with the perturbation theory remains good, if one compares shapes at the same distance to the critical point.

\section{Conclusions and outlook}

We investigated experimentally, numerically and theoretically the properties of minimal surfaces spanning two identical non-axisymmetric frames, with a particular focus on elliptic and cloverleaf frames. We  paid particular attention to the influence of the distance $h$ and the angle $\theta_0$ between the two frames on different properties of these surfaces. From a theoretical point of view, we propose a perturbative approach allowing to compute all the properties of the surface with good precision. The advantage of this approach in contrast to currently existing methods, such as proposed by Alimov at al \cite{alimov2021soap}, is that is can be easily generalised to a wide range of frame shapes, including surfaces contained between non-identical frames.

Our experimental, numerical and theoretical study first focused on the instability of the continuous minimal surface leading to a discontinuous minimal surface (Goldschmidt surface) at a critical height $h^\star$. We have shown experimentally and numerically that, surprisingly enough, $h^\star$ depends only slightly on the angle $\theta_0$. Perturbative theory fails to predict this non-linear effect, and to our knowledge no currently available theory predicts this variation.

We then investigated in more detail the shapes of the minimal surfaces, showing systematically excellent agreement between experiments and simulations, both agreeing very well with the prediction by Alimov {\it et al.} \cite{alimov2021soap} for the case of $\theta_0=0$. Perturbation theory is found to capture these shapes very well far from the instability. Close to the instability, this agreement is greatly improved when shapes at the same relative distance to the critical height $h^\star$ are compared, rather than shapes at the same height $h$. We also discussed the existence of a particular curve on the surface, consisting of points whose normal is perpendicular to the normal of the frames, which generalises the neck commonly defined for the catenoid.

We then studied the mechanical stresses (force and torque) exerted by the minimal surface on the frames. Since we consider surface tension to be constant, these physical quantities have a purely geometrical interpretation. We show that they present non-trivial behavior, including a discontinuity of both the force and the torque at $h=0$, as well as a non-monotonic character of the force for $\theta_0\neq 0$.  Experiments, Surface Evolver simulations, perturbation theory and the Alimov {\it et al.} method \cite{alimov2021soap} are in very good agreement.

Beyond the fundamental interest of this work, these results could pave the way for tools for future investigations. One of the most intriguing property of the non-axisymmetric minimal surfaces is its capacity to transmit not only normal stresses, but also a torque, despite being liquid. They could therefore be used to transmit or measure very small torques or normal forces. This could be used, for example, to investigate the visco-elastic properties of soap films. 
Since perturbation theory predicts the torque and normal force well, both can be determined unambiguously. By adjusting the frame geometry, the measuring range of the two quantities could be adapted.

We concentrated here on the analysis of a specific choice of shapes (ellipses and clovers), with both frames being identical, parallel and rotated around the central axis. In future work it will be interesting to investigate arbitrary shapes and more general rotations, including the influence of the frames not being parallel. It will also be interesting to extend this work to surfaces in which surface tension depends on deformation, simulating an elastic response relevant in material design \cite{mb} or for the description of biological membranes.

Last but not least, while perturbation theory captures well a wide range of system properties, we have seen that it neglects non-linear effects which become important close to the instability. It would therefore be important to develop more accurate theoretical tools, for example in generalising the approach proposed by Alimov {\it et al.} \cite{alimov2021soap} to more general surface shapes.

\begin{acknowledgments}
The authors wish to thank S. Boukhris, T. Boutfol, V. Grimaud, L. Louboutin and O. Ruelle for their participation in the experiments. We thank R. Bollache and J. Dijoux for their help with the experimental set-up and the 3D-printing. We thank S. Kaufmann, A. Hourlier-Fargette and M. Jouanlanne for numerous discussions and help with the soap formulations. We also thank G. Ginot for support on Surface Evolver questions and P. K\'ekicheff for sharing his lab space. We acknowledge financial support from an ERC Consolidator Grant (agreement 819511 - METAFOAM).
\end{acknowledgments}

\section*{Author Contributions}

Friedrich Walzel: Methodology, Validation, Data curation, Formal analysis, Visualization, Investigation, Writing - Original Draft, Writing - Review and Editing.  Alice Requier: Methodology, Validation, Data curation, Formal analysis, Visualization, Investigation, Writing - Original Draft, Writing - Review and Editing. Kevin Boschi: Investigation, Data curation, Formal analysis. Jean Farago: Conceptualization, Methodology, Formal analysis, Writing - Review and Editing. 
Philippe Fuchs:  Investigation, Data curation, Formal analysis. Fabrice Thalmann: Conceptualization, Methodology, Formal analysis, Writing - Review and Editing. Wiebke Drenckhan: Conceptualization, Methodology, Formal analysis, Writing - Original Draft, Writing - Review and Editing, Visualization, Supervision, Project administration, Funding acquisition. Pierre Muller: Conceptualization, Methodology, Formal analysis, Writing - Original Draft, Writing - Review and Editing, Visualization, Supervision, Project administration, Funding acquisition. Thierry Charitat: Conceptualization, Methodology, Formal analysis, Writing - Original Draft, Writing - Review and Editing, Visualization, Supervision, Project administration, Funding acquisition.

\appendix

\section{Materials and methods}
\subsection{Experiment}
\label{AppendixA01}

Minimal surfaces are studied using soap films held by 3D-printed frames with a set-up schematised in Fig. \ref{fig:setup_experiment}. The position of the lower frame remains fixed during the experiment, while the upper frame, attached to a vertical translation stage, can move at variable speed between controlled positions via a home written Labview program. The lower frame is fixed on a laboratory scale (METTLER TOLEDO, precision:  $0.1$ g) to measure the normal force $F_z$ acting between the frames. The ensemble is visualised from the side in front of a diffuse light source using a computer-controlled CCD camera with a spatial resolution of $50\ \mu m$. The frames are fabricated by a thermoplastic 3D printer Form 2 from Formlabs. The used printing method was stereolithography with a layer thickness of 0.025 mm. The deviation between the mathematical description, Eqs. \ref{eq:bound_el} and \ref{eq:bound_cl}, and the printed frame geometry is a maximum of $0.4\%$. The used soap solution is optimised for film stability: 500 ml of water, 22.5 ml of the dish washing liquid "Fairy", 7.5 ml of Glycerol and 3.0 g/L Jlube. All ingredients are mixed for 24 hours using a magnetic stirrer. The age of the solutions is maximally three month. The surface tension is measured using a "Kibron V2" tensiometre and determined to be $\gamma=25.7 \pm 0.1$ mN/m.

The minimal surface is produced by wetting the upper frame with the soap solution. Then the upper frame is moved downwards until it touches the lower frame. The upper frame is moved up by a few millimetres and the lower frame is rotated by $-\theta_0$. The distance $h$ between the frames is then slowly increased in small intervals $\Delta h$  and measurements are taken for the normal force and the shape. At every step it is controlled if the minimal surface still connects the two frames. If a Goldschmidt solution is observed, the critical height $h^\star$ is obtained by taking the average between the current $h$ and the previous height giving a stable shape.
Between each height change, at least five seconds wait time ensures static equilibrium of each shape.  We improve the precision on $h^\star$ by decreasing $\Delta h$ in its vicinity and by repeating measurements.\\ 
The images are treated by a home-made python code, which uses light intensity gradients to determine the distance $h$ between the frames and the projected contour of the minimal surface.
Normal force and torque are not measured simultaneously. The torque $\Gamma_z$ is measured with a Discovery HR-3 hybrid rheometre holding the same frames as in the normal force measurements. The same equilibrium protocol as for the normal force measurement is used. The precision in the torque is 5 nNm. The uncertainties in the measured quantities are mainly influenced by the differences between the mathematical description of the contours and the actual shape of the frames, by imperfect alignment of the frames (centering and parallelism), the resolution and sharpness of the camera. Additional uncertainties are related to the precision of the scale and the rheometre.

\subsection{Surface Evolver simulations}
\label{AppendixA02}

Surface Evolver is an open source Finite Element program which represents a surface via vertices, edges and facets \cite{evolver}. Surface Evolver minimises the total energy by moving vertices of a defined shape in the oposite direction of the energy gradient. 
The total energy is in our case proportional to the total area $A$, defined here as the sum of the facets areas, since only a constant surface tension is considered. Vertices on the frame stay fixed at the position defined by the Eqs. (\ref{eq:bound_el}) or (\ref{eq:bound_cl}).

Our simulation procedure is similar to the procedure in the experiment. The height $h$ between the two frames is increased in small steps $\Delta h$ until the surface becomes unstable.

After each change of height the surface is relaxed by moving the vertices until the relative energy change is smaller than $10^{-10}$ after $100$ such movements. The mesh is then optimised and the process is repeated until the relative  change is again smaller than $10^{-10}$ after $100$ relaxations. Iterations between these two steps stop if the convergence criterion is met twice in a row. To avoid that the system is trapped in a local minimum, all vertices are randomly moved by a small distance ("jiggled") at least twice during the relaxation process.

\begin{figure}[h!]
\includegraphics[width=0.75\textwidth]{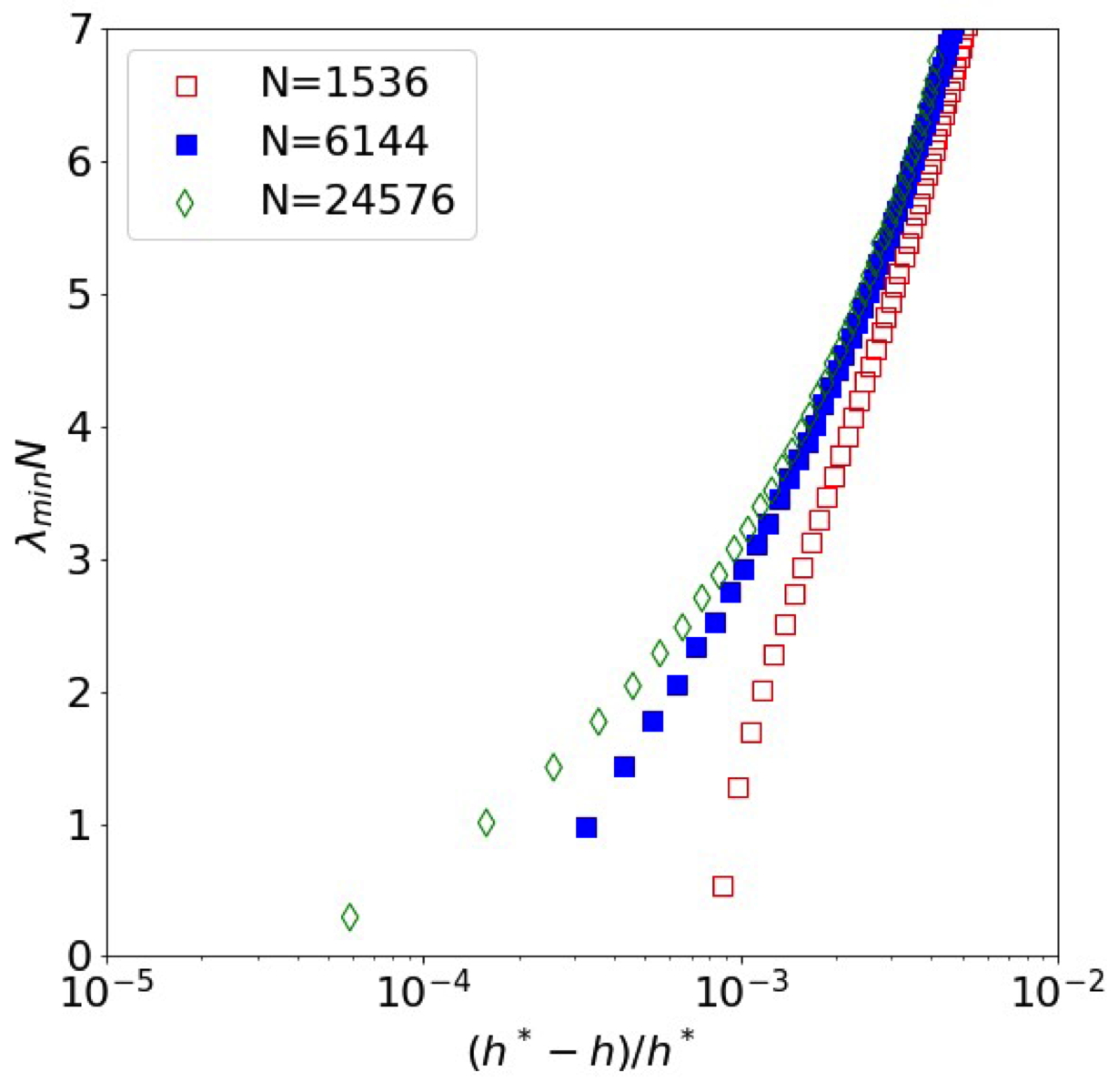}
\caption{Smallest eigenvalue $\lambda_{\rm min}$ (multiplied by the number $N$ of facets) of the Hessian matrix $ \underline{ \underline{H}}$  of the total energy in Surface Evolver for different numbers of facets $N$ in the case of the axisymmetric catenoid, as a function of the reduced height $(h^\star-h)/h^\star$, where $h^\star$ is the critical height of the axisymmetric catenoid. As the value of the latter is known exactly, the error for other simulations can be estimated depending on the number of facets used in this graph.}
\label{fig:eigenvalues_catenoid}
\end{figure}

To obtain the best precision in the critical height $h^\star$ we use the Hessian matrix $ \underline{ \underline{H}}$ of the area functional $A$ since the Taylor development of $A$ at fixed $h$ and $\theta_0$ up to the second order is given by
\begin{equation}
\label{eq:hessian_SE}
    A({\bf X}+\delta{\bf X})=A({\bf X})+{\bf\nabla}A\cdot{\delta\bf X}+\frac{1}{2} \delta{\bf X}^T\underline{\underline{H}} \delta{\bf X}.
\end{equation}
The free coordinates ${\bf X}$ are defined by the number of facets and their ability to change $A$ by a small perturbation of the coordinates of the vertices. If $A$ represents an extremal surface then ${\bf\nabla}A =0$. For it to be a minimum (and hence physically stable) all eigenvalues $\lambda$ of $ \underline{ \underline{H}}$ have to be strictly positive.
In evaluating the $\lambda$s, vertices are restricted to move along a direction normal to the local tangent plane of the surface, as tangent displacements are marginal, akin to redefining the surface mesh. The smallest eigenvalue $\lambda_{min}$ at the critical point is zero. Extrapolating the evolution of the smallest eigenvalue with increasing height of the last stable shapes gives a very good approximation of $h^\star$. Figure \ref{fig:eigenvalues_catenoid}  shows this for a simple catenoid for which $h^\star$ is known exactly. One observes that with increasing number of facets $N$ the relative error for the obtained critical height is converging towards zero. It also shows that the relative error is already small ($<0.04\%$) for a relatively small number of facets ($N=1536$). While there is a certain flexibility as far as defining the $\delta \bf X$ degrees of freedom, the moment where the Hessian becomes singular should be independent of such choice, up to numerical errors. This validates the general procedure and provides at the same time an estimation of the precision of the simulation.

The normal force $F_z$ and the torque $\Gamma_z$ on the axis passing by the two frame centers are related via

\begin{equation}
\label{eq:force_torque}
\gamma\dd A=\Gamma_z(\theta_0,h) \dd\theta_0 + F_z(\theta_0,h)\dd h.
\end{equation}

The force and the torque applied on the frame are the same along the surface. The derivatives $\dd A$, $\dd\theta_0$ and $\dd h$ are approximated by finite differences of two simulated surfaces with a small change in $h$ or $\theta_0$. The precision of these calculations depends strongly on the precision of the total area minimum, which, in turn, depends on the number of facets and the iteration process.

\section{Perturbation theory}
Here we present a perturbation approach to approximate minimal surfaces close to the catenoid. The two frames can therefore be of different shape, in contrast to the  model presented by Alimov \textit{et al.} \cite{alimov2021soap}.
\\The computation of the perturbation theory concerning the non-axisymmetric minimal surface consists in solving the minimal surface differential equation, corresponding to a vanishing mean curvature. We start by recalling the form of this partial differential equation in cylindrical coordinates, then we present in the general case the perturbative scheme we have developed to solve this equation and we discuss the conditions of the existence of a surface. Finally we show how the perturbative approach allows us to calculate the normal force and the torque exerted by the minimal surface on the contours.

\section{Minimal surface differential equation}
\label{appendixA}

The computation of the perturbative theory concerning the non-axisymmetric minimal surface consists in solving the minimal surface differential equation, in polar coordinates, corresponding to a vanishing mean curvature $H$. To do that, the expression of $H$ has to be determined. 

First, the coordinates of the vector position $\mathbf{r}$ are defined as follows
\begin{eqnarray}
&&x\left(\theta, z\right)=r\left(\theta, z\right)\cos{\theta}\\
&&y\left(\theta, z\right)=r\left(\theta, z\right)\sin{\theta}\\
&&z\left(\theta, z\right)=z.
\end{eqnarray}

Using the usual notations 
\begin{eqnarray}
&&{\bf r}_i\left(\theta, z\right)=\frac{\partial{\bf r}}{\partial i}\\ 
&&{\bf r}_{ij}\left(\theta, z\right)=\frac{\partial^2{\bf r}}{\partial i\partial j},
\end{eqnarray}
with $i,j=\theta,z$, the normal vector to the surface can be calculated thanks to its definition ${\bf n}={\bf r}_z\times{\bf r_\theta}/\left\|{\bf r}_z\times{\bf r_\theta}\right\|$

\begin{equation}
{\bf n}=\frac{1}{\left(r^2+r_\theta^2+r^2r_z^2\right)^{1/2}}\begin{pmatrix}
-r\cos{\theta} -r_\theta\sin{\theta}\\[3mm]
-r\sin{\theta}+r_\theta\cos{\theta} \\[3mm]
rr_z \\
\end{pmatrix}
\end{equation}

Then, the coefficients of the first fundamental form $\mathcal{F}_1$ are computed

\begin{eqnarray}
E&=&{\bf r}_z.{\bf r}_z=1+r_z^2~;~G={\bf r}_\theta.{\bf r}_\theta=r^2+r_\theta^2~;\\ 
F&=&{\bf r}_z.{\bf r}_\theta =r_\theta r_z,
\end{eqnarray}


as well as those of the second fundamental form $\mathcal{F}_2$

\begin{eqnarray}
L&=&{\bf r}_{zz}.{\bf n} =\frac{-rr_{zz}}{\left(r^2+r_\theta^2+r^2r_z^2\right)^{1/2}};\\
N&=&{\bf r}_{\theta\theta}.{\bf n}=\frac{r^2+2r_\theta^2-r r_{\theta\theta}}{\left(r^2+r_\theta^2+r^2r_z^2\right)^{1/2}};\\
M&=&{\bf r}_{z\theta}.{\bf n}=\frac{r_\theta r_z-r r_{z\theta}}{\left(r^2+r_\theta^2+r^2r_z^2\right)^{1/2}}.
\end{eqnarray}

So the expression of the mean $H$ and Gaussian $K$ curvature arises from these six coefficients
\begin{eqnarray}
\label{Hmeangene}
H &=& \frac 12\left(H_1+H_2\right) = \frac{EN+GL-2FM}{EG-F^2},\\
\label{Kmeangene}
K &=& H_1H_2 = \frac{LN - M^2}{EG-F^2}.
\end{eqnarray}
The two principal curvatures $H_1$ and $H_2$ are then given by:
\begin{eqnarray}
\label{H1gene}
H_1 &=& H + \sqrt{H^2-K},\\
\label{H2gene}
H_2 &=& H - \sqrt{H^2-K}.
\end{eqnarray}

The following partial differential equation results from this vanishing mean curvature problem defining minimal surfaces
\begin{eqnarray}
\nonumber
r_\theta^2+r\left[r\left(1+r_z^2\right) - r_{\theta\theta}\left(1+r_z^2\right) - r^2 r_{zz}\right. \\
\left.+ r_\theta\left(2 r_z r_{z\theta}-r_\theta r_{zz}\right)\right]=0.
\label{eq:equadiffgene_appendix}
\end{eqnarray}

If we suppose that the minimal surface considered is invariant by rotation (corresponding to $r_{\theta}=r_{\theta\theta}=r_{z\theta}=0$), we obtain $1+r_z^2 - r r_{zz} = 0$, which is the equation whose solution is nothing but the symmetric catenoid for $z \in [-h/2 ; h/2]$
\begin{equation}
\label{asymmetric_cat}
r_0\left(z\right) = a_c\cosh{\left(\frac{z}{a_c}\right)},
\end{equation}
where $a_c$ is solution of
\begin{equation}
\label{eq:neck_radius_appendix}
R = a_c\cosh{\left(\frac{h}{2a_c}\right)},
\end{equation}
which ensures the boundary condition $r\left(z=\pm h/2\right)=R$.

More generally, in the case asymmetric minimal surface spanning on circular frames of different radius $R_1$ and $R_2$, we have to introduce a new constant $C_c$ as
\begin{equation}
\label{eq:shape_appendix}
r_0\left(z\right) = a_c\cosh{\left(\frac{z}{a_c}+C_c\right)}.
\end{equation}
$a_c$ and $C_c$ are now solutions of
\begin{eqnarray}
\label{eq:neck_radius_gene}
R_1 &=& a_c\cosh{\left(\frac{h}{2a_c}+C_c\right)},\\
R_2 &=& a_c\cosh{\left(-\frac{h}{2a_c}+C_c\right)}.
\end{eqnarray}
The existence of solutions is discussed in detail in \cite{Salkin}.

\section{Perturbative theory}
\label{appendixB}
We want to investigate a special group of minimal surfaces, spanning two {\bf non-axisymmetric} closed frames, here not necessarily identical, which are contained in two parallel planes. The planar boundaries ${\cal C}_\pm$ ($+$ and $-$ denoting the upper and lower frame, respectively) are centered on the $z$ axis and given in polar coordinates $r_{\mathcal{C}_\pm}\left(\theta\right)$. They are separated by a distance $h$.

We assume that it is possible to do a Fourier decomposition
\begin{equation}
r_{\mathcal{C}_\pm}\left(\theta\right)=R_{\rm mean}^\pm\left(1 + \sum_{k=2}^{+\infty} a_k^\pm\cos{\left(k\theta\right)} + b_k^\pm \sin{\left(k\theta\right)}\right),
\label{fourier_contour_gene}
\end{equation}
where $R_{\rm mean}^\pm$, $a_k^\pm$ and $b_k^\pm$ are respectively the mean radius and the Fourier coefficients of $\mathcal{C}_\pm$.

\subsection{General perturbative approach}
\label{PT-calcul}

The idea of the perturbative approach is to solve Eq. (\ref{eq:equadiffgene_appendix}) considering surfaces close to the asymmetric catenoid. Therefore we take Eq. (\ref{asymmetric_cat}) and write the minimal surface shape in the form of
\begin{equation}
r\left(\theta, z\right)=a\cosh{\left(\frac{z}{a}+C\right)}\left(1+f\left(\theta, z \right)\right),
\label{eq:r_appendix}
\end{equation}
and consider $f(\theta, z)$ as a perturbative term.

Rewriting the equation \ref{eq:equadiffgene_appendix} on $f$ and restricting it to the first order in $f$, we end up with a linear differential equation in the form
\begin{equation}
\label{equadifordre1_appendix}
2 f\left(\theta, z\right) + \cosh^2{\left(\frac{z}{a}+C\right)} \left(f_{\theta\theta}\left(\theta, z\right) +a^2 f_{zz}\left(\theta, z\right)\right)=0.
\end{equation}

To solve this equation, let us decompose $f$ in Fourier series
\begin{equation}
\label{fouriergene}
f\left(\theta, z\right) = \sum_{m=1}^{+\infty}\alpha_m\left(z\right)\cos{\left(m\theta\right)}+\beta_m\left(z\right)\sin{\left(m\theta\right)}.
\end{equation}
Introducing this expansion in Eq. (\ref{equadifordre1_appendix}), we obtain the following equations for the different modes
\begin{widetext}
\begin{eqnarray}
\left(2 - m^2\cosh^2{\left(\frac{z}{a}+C\right)}\right)\alpha_m\left(z\right) + a^2\cosh^2{\left(\frac{z}{a}+C\right)} \alpha_m^{\prime\prime}\left(z\right)=0,\\
\left(2 - m^2\cosh^2{\left(\frac{z}{a}+C\right)}\right)\beta_m\left(z\right) + a^2\cosh^2{\left(\frac{z}{a}+C\right)} \beta_m^{\prime\prime}\left(z\right)=0,
\label{equationalphabeta}
\end{eqnarray}
\end{widetext}

which can be solved to a given order $m_0$, with boundary conditions on $\alpha_{m}$ and $\beta_{m}$ which are given the geometry of the frames
\begin{eqnarray}
\label{bound}
r\left(\theta,h/2\right)=r_{\rm \mathcal{C},+}\left(\theta\right),\\
r\left(\theta,-h/2\right)=r_{\rm \mathcal{C},-}\left(\theta\right),
\end{eqnarray}
leading to, using the boundary conditions above as well as Eqs. (\ref{eq:r_appendix}), (\ref{fouriergene}) and (\ref{rc})
\begin{eqnarray}
\label{equationCL1el}
a\cosh{\left(C+\frac{h}{2a}\right)} &=& R_{\rm mean}^+,\\
\label{equationCL2}
a\cosh{\left(C-\frac{h}{2a}\right)} &=& R_{\rm mean}^-,\\
\label{boundaryalphael}
\alpha_k\left(\pm h/2\right) &=& a_k^\pm,\\
\label{boundarybetael}
\beta_k\left(\pm h/2\right) &=& b_k^\pm.
\end{eqnarray}

For the different geometries investigated below, the differential equations \ref{equationalphabeta} were solved using Python (function solve\_bvp from scipy package).

Finally, It is interesting to note that for high order $k\gg 1$, Eqs. (\ref{equationalphabeta}) can be simplified using $k^2\cosh^2{z/}\gg 2$ leading to much simpler homogeneous linear second order equations
\begin{eqnarray}
- k^2\alpha_k\left(z\right) + a^2 \alpha_k^{\prime\prime}\left(z\right)=0,\\
 - k^2\beta_k\left(z\right) + a^2 \beta_k^{\prime\prime}\left(z\right)=0.
\label{equadifbeta_klarge}
\end{eqnarray}
The  solutions for these approximated differential equations are then
\begin{eqnarray}
\nonumber
\tilde{\alpha}_k\left(z\right)&=&\frac{a^-_k\sinh{\left(\frac ka\left(\frac h2 - z\right)\right)} + a^+_k\sinh{\left(\frac ka\left(\frac h2 + z\right)\right)}}{\sinh{\left(\frac{kh}{a}\right)}}\\
\\
\nonumber
\tilde{\beta}_k\left(z\right)&=&\frac{b^-_k\sinh{\left(\frac ka\left(\frac h2 - z\right)\right)} + b^+_k\sinh{\left(\frac ka\left(\frac h2 + z\right)\right)}}{\sinh{\left(\frac{kh}{a}\right)}}.
\label{equadifbeta_klarg_solution}
\end{eqnarray}

\subsection{Case of identical frames}

When the two frames share the same shape defined by
\begin{equation}
\label{eq:contour_frame}
r_{\mathcal{C}}\left(\theta\right)=R_{\rm mean}\left(1 + \sum_{k=2}^{+\infty} a_k\cos{\left(k\theta\right)} + b_k \sin{\left(k\theta\right)}\right),
\end{equation}
we can introduce the angle $\theta_0$ between the frame. The lower and upper boundary are thus given by
\begin{eqnarray}
\label{rc}
r_{\rm \mathcal{C},\pm}\left(\theta\right)&=&r_{\rm \mathcal{C}}\left(\theta \pm \frac{\theta_0}{2}\right)
\end{eqnarray}
and we can express the Fourier coefficients $a_k^\pm$ and $b_k^\pm$ as
\begin{eqnarray}
\label{fourier_syma}
&&a_k^\pm = a_k\cos{\left(k\frac{\theta_0}{2}\right)}\mp b_k\sin{\left(k\frac{\theta_0}{2}\right)},\\
\label{fourier_symb}
&&b_k^\pm = \pm a_k\sin{\left(k\frac{\theta_0}{2}\right)}+b_k\cos{\left(k\frac{\theta_0}{2}\right)}.
\end{eqnarray}

The boundary conditions leads to
\begin{eqnarray}
\label{equationCL1id}
&&a\cosh{\left(\frac{h}{2a}\right)} = R_{\rm mean}\ {\rm and}\ C=0,\\
\label{boundaryalphaid}
&&\alpha_k\left(\pm h/2\right) = a_k\cos{\left(k\frac{\theta_0}{2}\right)}\mp b_k\sin{\left(k\frac{\theta_0}{2}\right)},\\
\label{boundarybetaid}
&&\beta_k\left(\pm h/2\right) = \pm a_k\sin{\left(k\frac{\theta_0}{2}\right)}+b_k\cos{\left(k\frac{\theta_0}{2}\right)}.
\end{eqnarray}

We have first to find $a$ by solving the first equation that corresponds to the usual equation for a catenoid, but with the mean radius $R_{\rm mean}$. In agreement with the catenoid's theory, Eq. (\ref{equationCL1id}) has two solutions $a_+$ and $a_-$ when $h < h^*_{\rm theo} = 1.33 R_{\rm mean}$. In the perturbative theory, the critical height is therefore proportional to the mean radius, whatever the shape of the frame, and whatever the angle $\theta_0$ between the frames.

\subsubsection{Case of minimal surface supported by ellipses}
\label{PT-ellipses}

\begin{figure}[h!]
    \centering
    \includegraphics[width=0.49\textwidth]{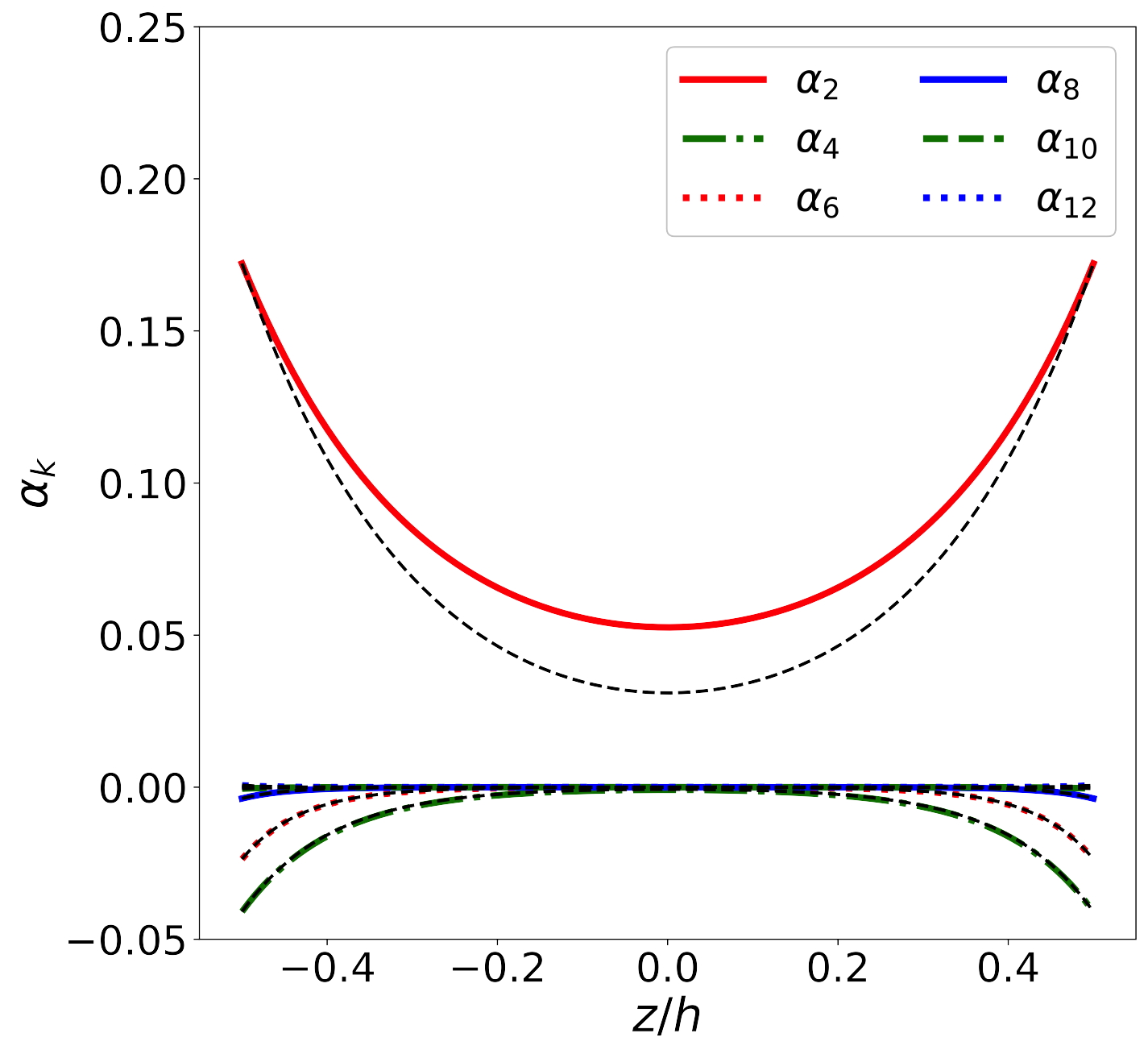}
    \includegraphics[width=0.49\textwidth]{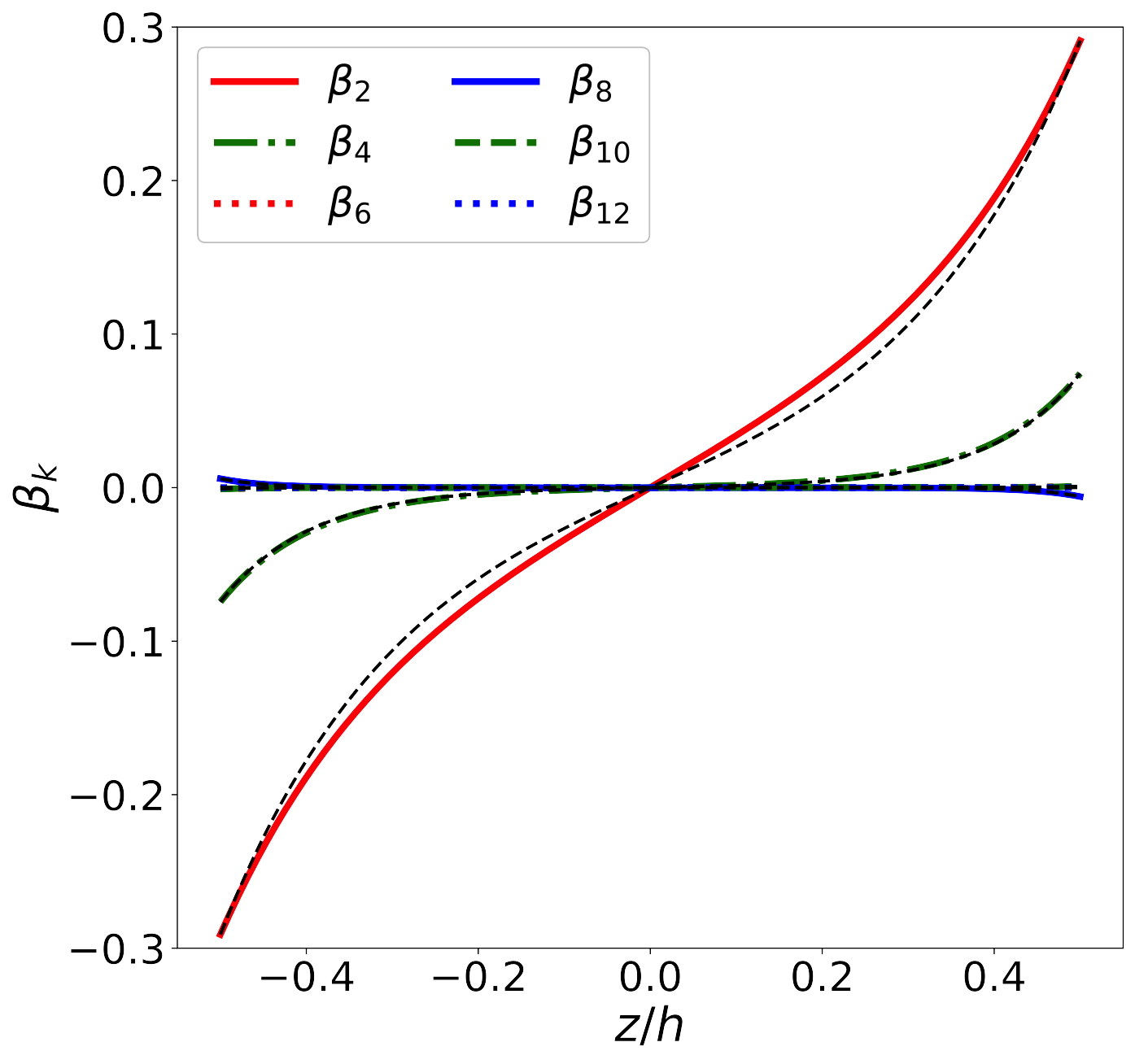}
    \caption{Solution of the differential equations (\ref{equationalphabeta}) in the case of elliptic frames ($e=0.97$, $R_{\rm mean}=1$, $h/R_{\rm mean}=1.4$ and $\theta_0=60^\circ$) for $\alpha_k\left(z\right)$ (a) and $\beta_k\left(z\right)$ (b) for $k\leq 12$. The black dashed lines are the approximation for $k\gg 1$, Eq. (\ref{equadifbeta_klarg_solution}).}
    \label{fig:alpha_beta_ellipse}
\end{figure}

A special case of the description developed above would be a minimal surface supported by ellipses of eccentricity $e$,  major axis $2R_{\rm M}$, and minor axis $2R_{\rm m}$, with $e^2=1-(R_{\rm m}/R_{\rm M})^2$ and $R_{\rm M}=R_{\rm m}/\sqrt{1-e^2}$. The polar equation of the ellipse with long axis along the $Ox$ direction is given by

\begin{equation}
r_{\rm e}\left(\theta\right)=\frac{R_{\rm m}}{\sqrt{1-e^2\cos^2{\left(\theta\right)}}}.
\end{equation}
We will also need the perimeter of the ellipse that is given by $P=4E[e^2]R_{\rm M}$.\\

The Fourier transform of the ellipse gives  $r_{\rm e}\left(\theta\right)=R_{\rm mean}\left(1+\sum_{m=2}^{+\infty} a_{\rm m}\left(e\right)\cos{\left(m\theta\right)}\right)$. For odd $m$, $a_m(e)$ is zero due to the symmetry of the ellipse about the $Ox$ axis. The first Fourier coefficients are equal to
\begin{widetext}
\begin{eqnarray}
R_{\rm mean}\left(e\right) &=& \frac 2\pi K\left[e^2\right]R_{\rm m},\\
a_2\left(e\right) &=& \frac{2}{e^2}\left(2 - e^2 - \frac{E\left[e^2\right]}{K\left[e^2\right]}\right),\\
a_4\left(e\right) &=& \frac{2}{3e^4}\left(16 - 16e^2 +3e^4 - 8\frac{\left(2-3e^2+e^4\right)}{\left(1-e^2\right)}\frac{E\left[e^2\right]}{K\left[e^2\right]}\right),\\
a_6\left(e\right) &=& \frac{2}{15e^6}\left(256 - 384e^2 + 158e^4 - 15e^6 - 2\left(128-128e^2+23e^4\right)\frac{E\left[e^2\right]}{K\left[e^2\right]}\right),
\end{eqnarray}
\end{widetext}
where $K[x], E[x]$ are the complete elliptic integral of the first and second kind. We have computed the Fourier coefficient of the ellipse up to order 12.

Knowing the $a_k$ coefficients, is it possible to solve differential Eqs. (\ref{equationalphabeta}) on $\alpha_k$ and $\beta_k$ with boundary conditions (\ref{equationCL1id}), (\ref{boundaryalphaid}) and (\ref{boundarybetaid}). We have plotted the first 12 functions $\alpha_k$ and $\beta_k$ on Fig.\ref{fig:alpha_beta_ellipse} for elliptic frames ($e=0.97$, $h=0.5$ ($R_{\rm M}=1$, $R_{\rm m}=0.24$) and $\theta_0=0.3\pi/2$). We have also plotted the approximated solutions given by Eqs.( \ref{equadifbeta_klarg_solution}) and that can be express as
\begin{eqnarray}
&\tilde{\alpha}_k\left(z\right) = \frac{\cosh\left(\frac{kz}{a}\right)}{\cosh\left(\frac{kh}{2a}\right)}\cos{\left(\frac{k\theta_0}{2}\right)} a_k,\\
&\tilde{\beta}_k\left(z\right) = \frac{\sinh\left(\frac{kz}{a}\right)}{\sinh\left(\frac{kh}{2a}\right)}\sin\left(\frac{k\theta_0}{2}\right) b_k.
\label{equadifbeta_klarg_solution_elliptic}
\end{eqnarray}

The results for the example of two elliptic frames with $e=0.97$, $h=0.5$ and $\theta_0=0.3\pi/2$ are shown in Fig. \ref{fig:alpha_beta_ellipse}. One observes easily how the coefficients converge for higher $k$ towards Eqs. (\ref{equadifbeta_klarg_solution}) given by the black lines.

\begin{figure}[h!]
\centering
{\label{fig:shape}\includegraphics[width=0.75\textwidth]{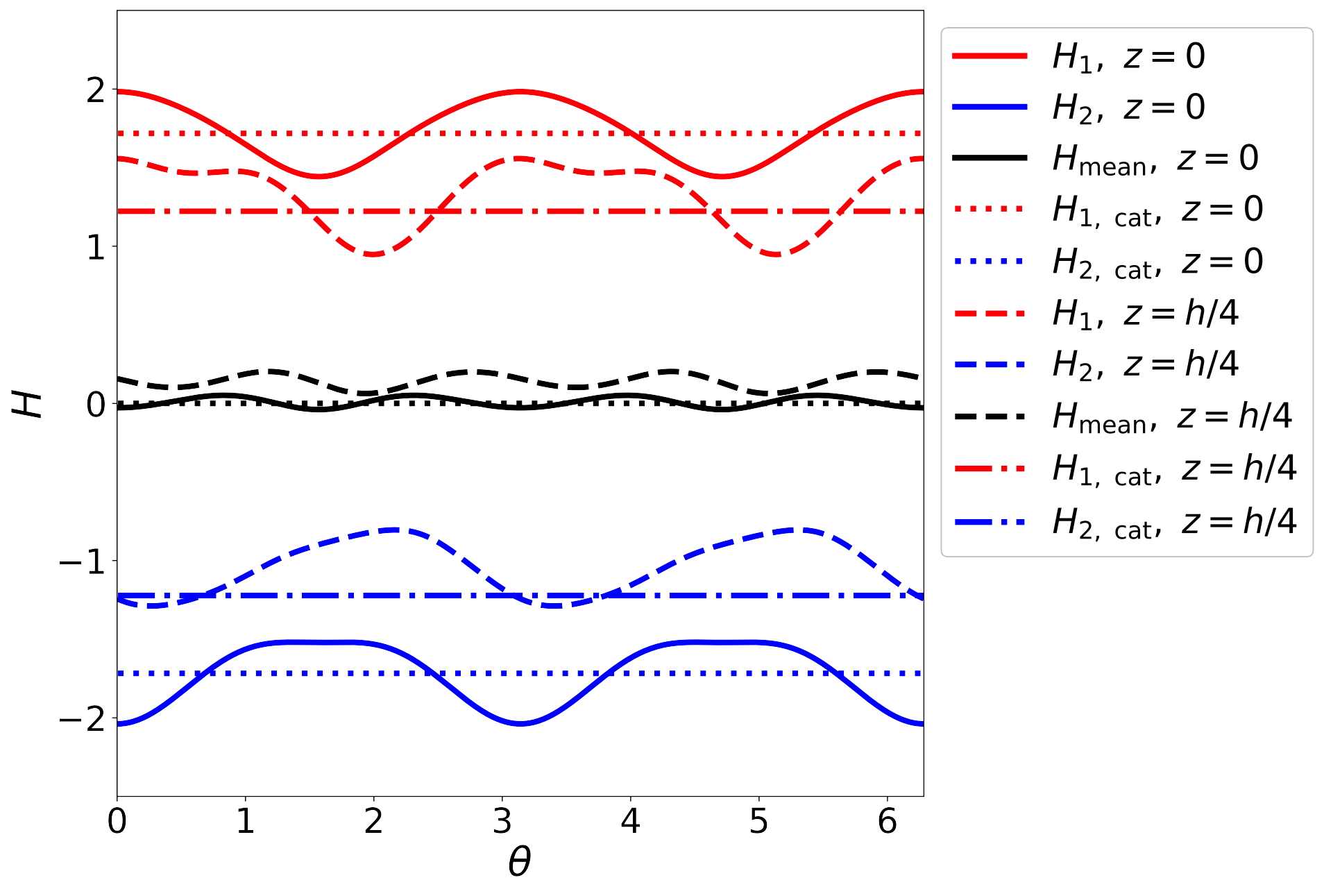}}
\caption{Principal curvatures $H_1$ (red curves) and $H_2$ (blue curves) and mean curvature (black curves) $H_{\rm mean}$  in the case of elliptic frames ($e=0.97$, $R_{\rm mean}=1$, $h/R_{\rm mean}=1.4$ and $\theta_0=60^\circ$) for two horizontal planes $z=0$ (solid lines) and $z=h/4$ (dashed lines). Dashed-dotted line gives the principal curvatures for the catendoid of the same mean radius for comparison.}
\label{fig:curvature}
\end{figure}

A good test of the perturbation theory is to calculate the mean curvature $H_{\rm mean}$ using Eq. (\ref{Hmeangene}). We have plotted this curvature as a function of $\theta$ for different horizontal planes $z=0$ and $z=h/4$ in Fig. \ref{fig:curvature}, with the principal curvatures $H_1$ and $H_2$.

\subsubsection{Clover contours}
\label{random}

We have also considered a three-leaves clover frames parametrised as follows, for $\theta \in [0;2\pi]$:

\begin{eqnarray}
\label{xc}
x_{\rm cl}\left(\theta\right) = R_{\rm mean}\left(1-\varepsilon\cos{3\theta}\right)\cos{\theta} \\
y_{\rm cl}\left(\theta\right) = R_{\rm mean}\left(1-\varepsilon\cos{3\theta}\right)\sin{\theta}.
\label{yc}
\end{eqnarray}

Numerical and theoretical shapes in the case of clover contours are reported on Figure 14.

\begin{figure}[!h]
    \centering
    \subfigure[]{\label{fig:trefle03}\includegraphics[width=0.18\textwidth]{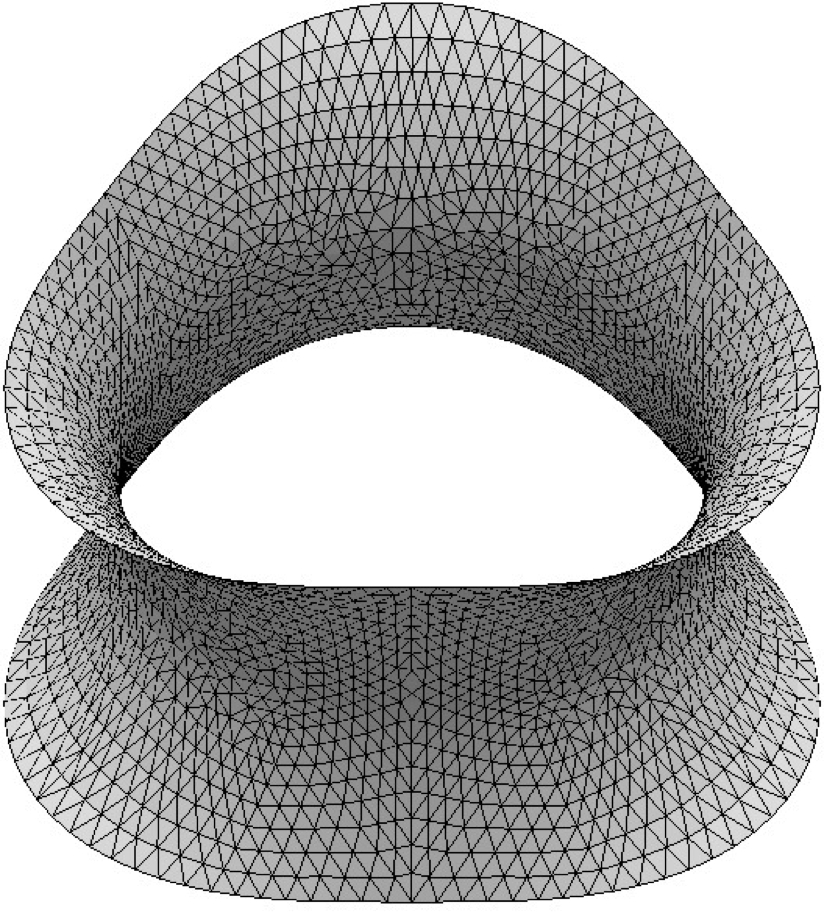}}
    \hspace{2cm}
    \subfigure[]{\label{fig:trefle01}\includegraphics[width=0.18\textwidth]{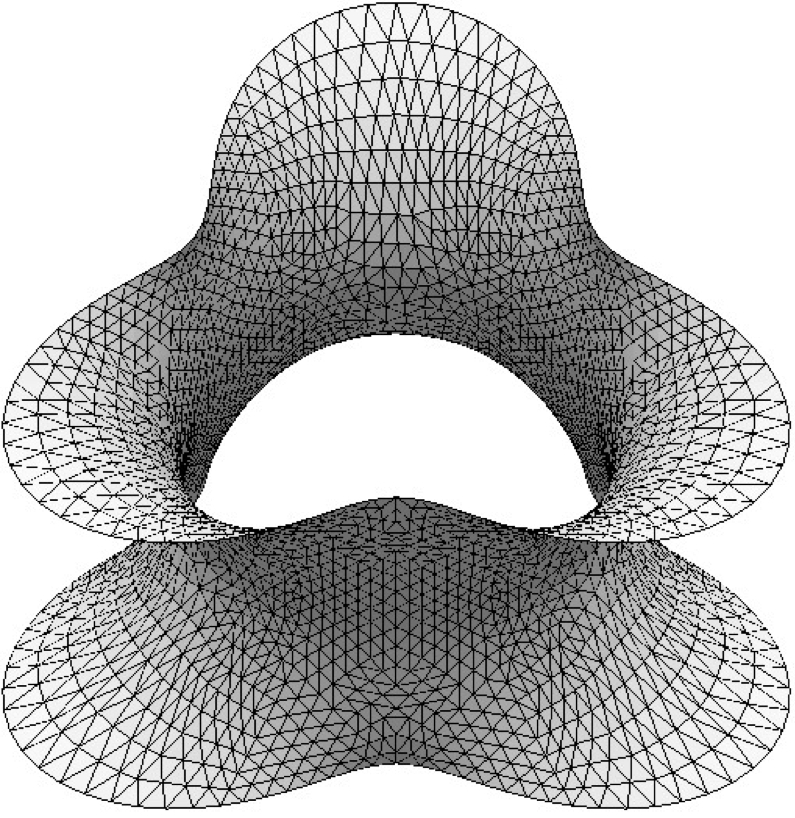}}

    \subfigure[]{\label{fig:trefle01}\includegraphics[width=0.18\textwidth]{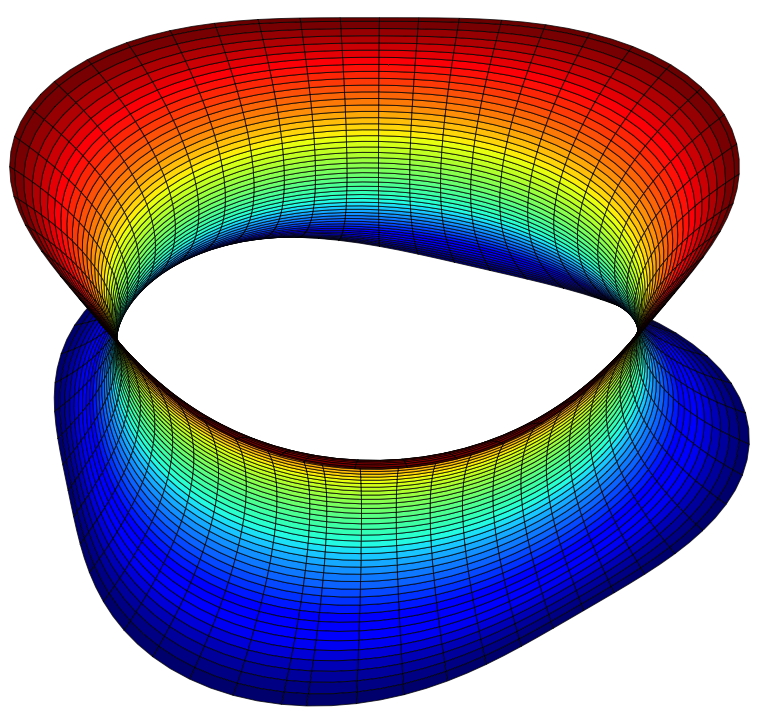}}
    \hspace{2cm}
    \subfigure[]{\label{fig:trefle01}\includegraphics[width=0.18\textwidth]{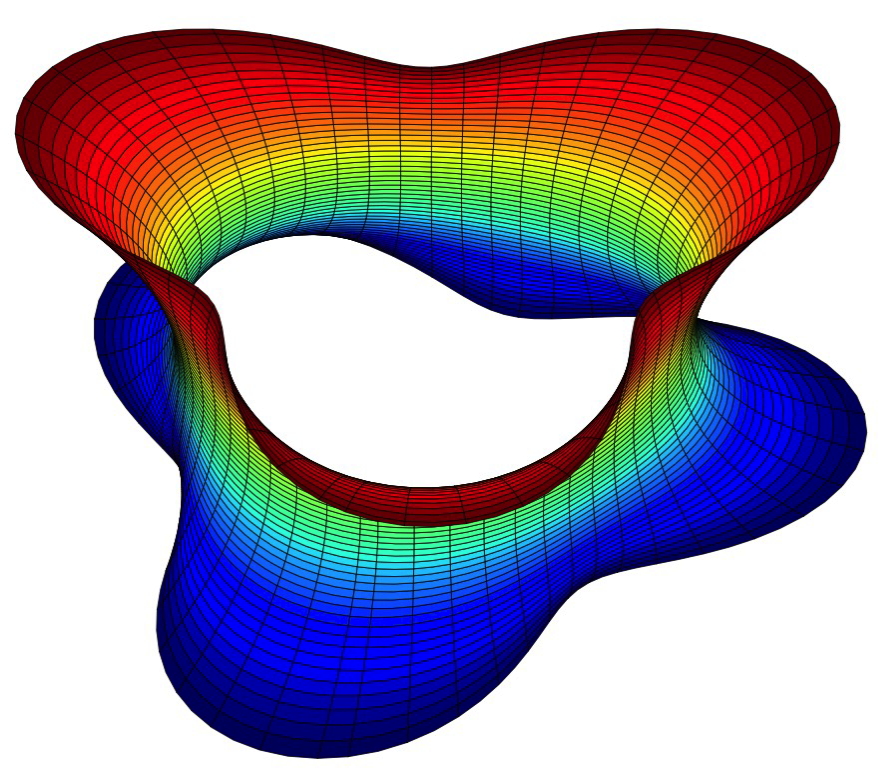}}
    \caption{Minimal surfaces supported by three-leaves clovers for $\varepsilon=0.1$ (Left) and $\varepsilon=0.3$ (Right): (a) and (b) numerical simulations (SE)  for $\theta_0=0$ and (c) and (d) perturbative theory for $\theta_0=20^\circ$.}
    \label{fig:trefles}
\end{figure}

\subsection{Non-identical frames}
\label{different-frames}

\begin{figure}[h!]
    \centering
    \includegraphics[width=0.49\textwidth]{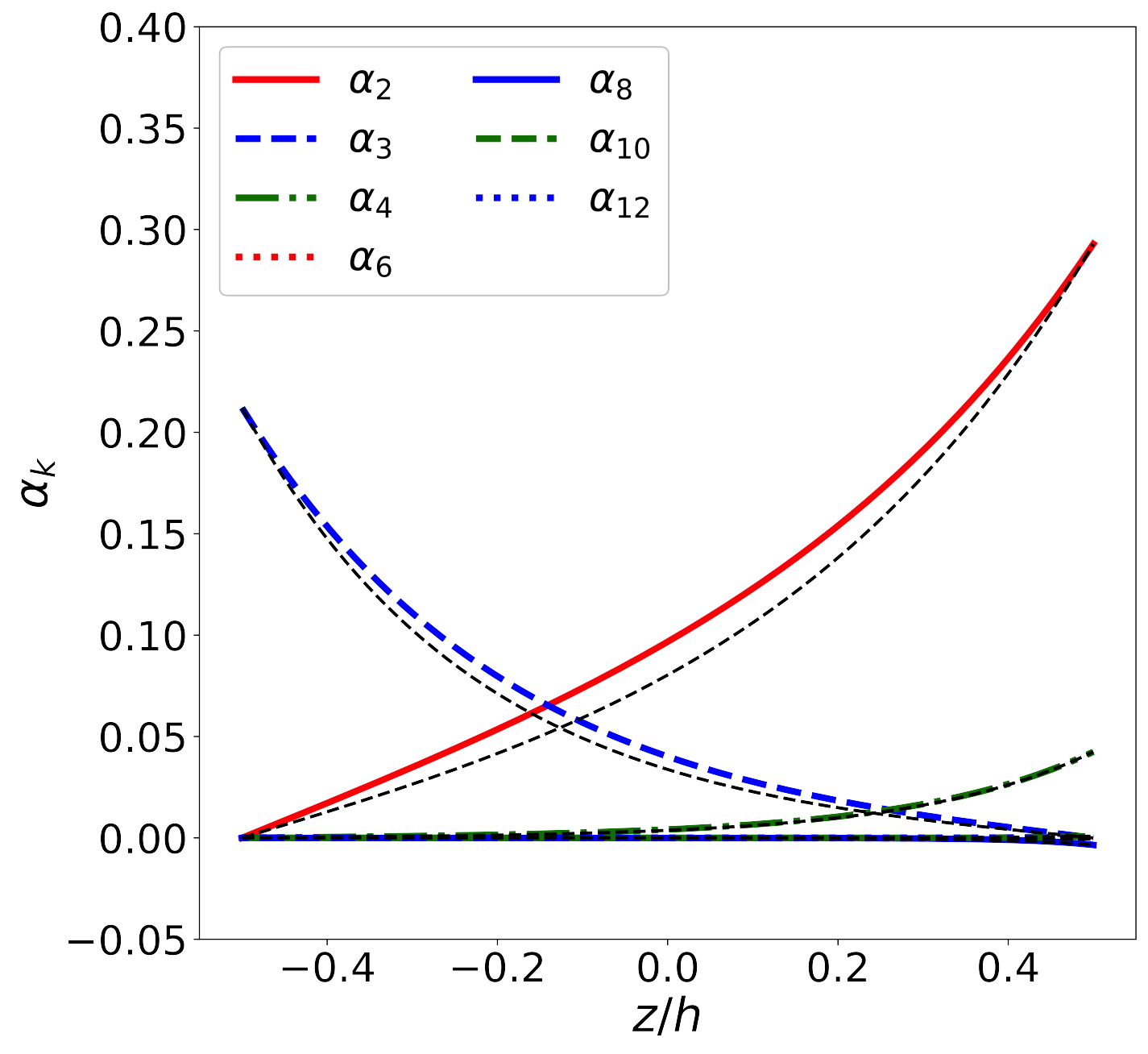}
    \includegraphics[width=0.49\textwidth]{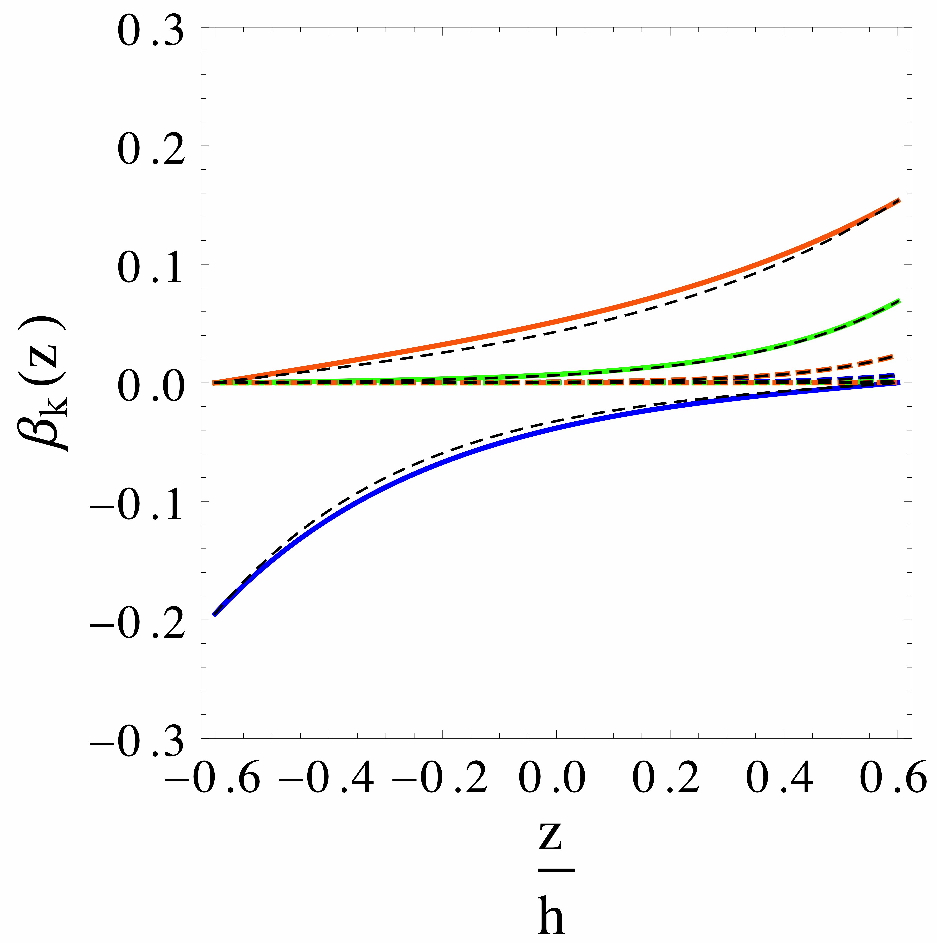}
    \caption{Solution of the differential equations (\ref{equationalphabeta}) for $\alpha_k\left(z\right)$ (a) and $\beta_k\left(z\right)$ (b) for $k\leq 12$, in the case of an upper elliptic frames ($e=0.866$, $R_{\rm M}=2.4$, $R_{\rm m}=1.2$,  $R_{\rm mean}^+=1.65$) and a lower three-leaves clover frame ($\varepsilon=0.3$, $R_{\rm mean}^-=1$) for $h=1.2$ and $\theta_0=30^\circ$. The black dashed lines are the approximation for $k\gg 1$, Eq. (\ref{equadifbeta_klarg_solution}).}
    \label{fig:alpha_beta_ellipse_asy}
\end{figure}

To test our approach in the case of two different frame, we used an ellipse ($e=0.866$, $R_{\rm mean}^+=1.2$) for the top frame and a three-leaves clover ($\epsilon=0.3$, $R_{\rm mean}=1$) for the lower one. In that case we have solved the differential Eqs. (\ref{equationalphabeta}) leading to the solution plotted on Fig. \ref{fig:alpha_beta_ellipse_asy}.

\begin{figure}[h!]
\includegraphics[width=0.49\textwidth]{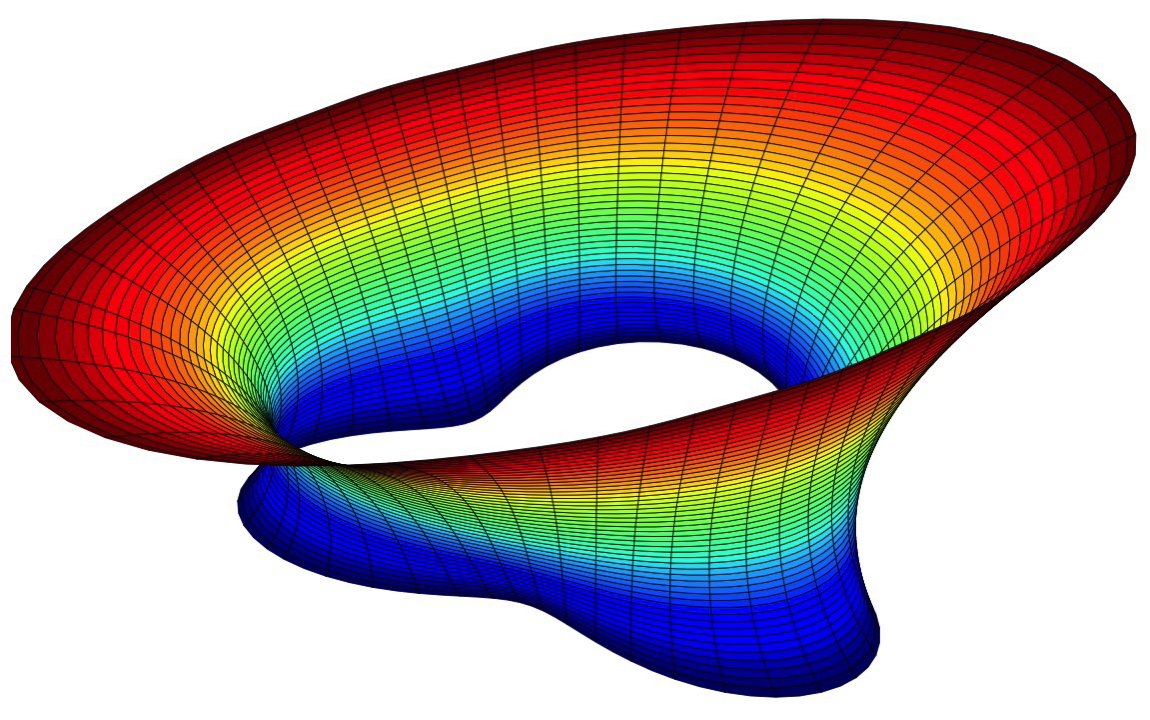}
\includegraphics[width=0.49\textwidth]{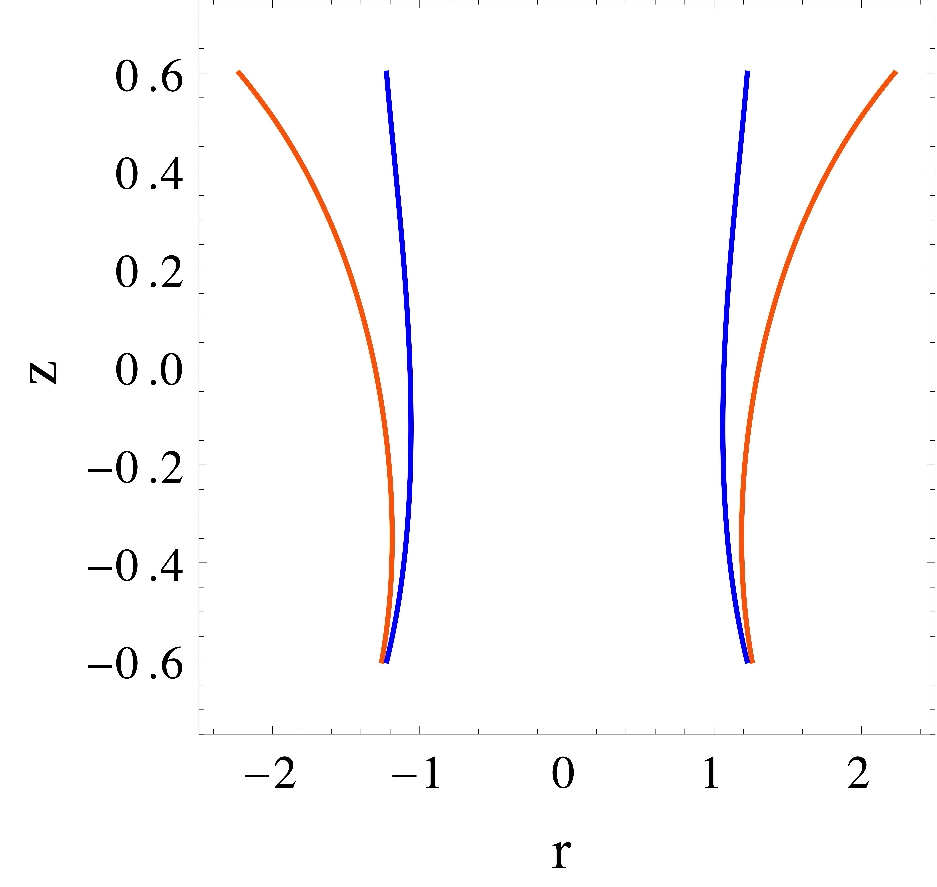}
\caption{Minimal surfaces in the case of an upper elliptic frame ($e=0.866$, $R_{\rm mean}^+=1.65$) and a lower three-leaves clover frame ($\varepsilon=0.3$, $R_{\rm mean}^-=1$) for $h=1.2$ and $\theta_0=30^\circ$: (Top): surface shape and (Bottom) profiles for $\theta_0=0^\circ$ (orange) and $\theta=90^\circ$ (blue).}
\label{fig:trefle-ellipse}
\end{figure}

We also reported the shape and profiles on Fig. \ref{fig:trefle-ellipse}.

\section{Force and torque predictions}
\label{appendixC}
\subsection{General relations}

The force and the torque applied by the minimal surface on the ellipses can also be computed using the perturbation theory. The elementary surface tension force acting on an element $\dd S$ is given by $\dd{\bf F}=2\gamma {\bf t}_e\times{\bf n}\dd S$ where ${bf n}$ is the vector normal to the surface and ${\bf t}_e$ is the tangent to the ellipse. We obtained for the total force acting on the ellipse~:

\begin{widetext}
\begin{eqnarray}
\frac{F_{{\mathcal{C}_\pm},x}}{2\gamma}&=&\int_{0}^{2\pi}\dd\theta\frac{-r\left(\theta, \pm h/2\right)r_z\left(\theta, \pm h/2\right)\left(r_{\mathcal{C}_\pm}^\prime\left(\theta\right)\sin{\theta}+r_{\mathcal{C}_\pm}\left(\theta\right)\cos{\theta}\right)}{\sqrt{r\left(\theta, \pm h/2\right)^2\left(1+r_z\left(\theta, \pm h/2\right)^2\right)+r_\theta\left(\theta, \pm h/2\right)^2}}\\
\frac{F_{{\mathcal{C}_\pm},y}}{2\gamma}&=&\int_{0}^{2\pi}\dd\theta\frac{r\left(\theta, \pm h/2\right)r_z\left(\theta, \pm h/2\right)\left(r_{\mathcal{C}_\pm}^\prime\left(\theta\right)\cos{\theta}-r_{\mathcal{C}_\pm}\left(\theta\right)\sin{\theta}\right)}{\sqrt{r\left(\theta, \pm h/2\right)^2\left(1+r_z\left(\theta, \pm h/2\right)^2\right)+r_\theta\left(\theta, \pm h/2\right)^2}}\\
\frac{F_{{\mathcal{C}_\pm},z}}{2\gamma}&=&\int_{0}^{2\pi}\dd\theta\frac{\left(r_\theta\left(\theta, \pm h/2\right)r_{\mathcal{C}_\pm}^\prime\left(\theta\right)+r\left(\theta, \pm h/2\right)r_{\mathcal{C}_\pm}\left(\theta\right)\right)}{\sqrt{r\left(\theta, \pm h/2\right)^2\left(1+r_z\left(\theta, \pm h/2\right)^2\right)+r_\theta\left(\theta, \pm h/2\right)^2}}.
\end{eqnarray}
\end{widetext}

Since the perturbative theory give not an exact minimale surface, the force equilibrium is not perfectly fulfilled and the calculated force is slightly depending on the integration contour.

It is also possible to calculate the torque using $\dd{\Gamma}={\bf r_e}\wedge{\bf\dd F}$~:

\begin{eqnarray}
\Gamma_z &=& 2\gamma\int_{0}^{2\pi}\dd\theta r_e\left(\theta\right)\left(\cos{\theta}\dd F_x - \sin{\theta}\dd F_y\right).
\end{eqnarray}

\subsection{Case $h=0$}
In the case $h=0$ the minimal surface is perfectly known. If $\theta_0=0$, then it is an infinitely thin cylinder generated by the frame. In that case the force is simply proportional to the perimeter of the ellipse $F_z=2\gamma P$.

\begin{figure}[h!]
\centering
\includegraphics[width=0.49\textwidth]{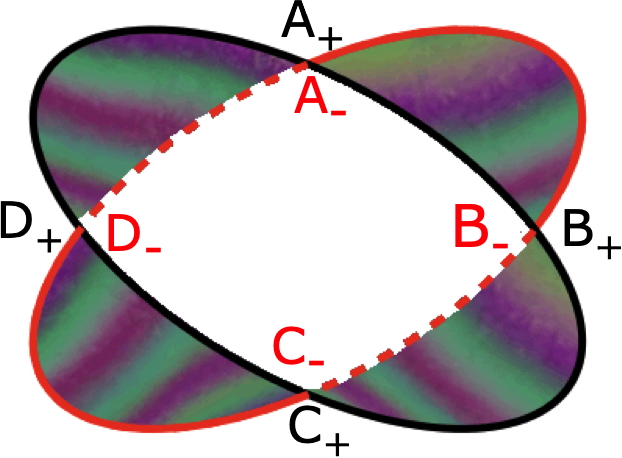}  \caption{In the case $h=0$ the soap film consist in the symmetric difference ${\cal S}_+\Delta{\cal S}_-$ of the surface of the two frames as shown in the elliptic case. The points of intersection between the two frames for $h=0$ become the points $A_+$, $B_+$, $C_+$ and $D_+$ (respectively  $A_-$, $B_-$, $C_-$ and $D_-$) for the bottom (respectively top) frame when $h\neq 0$.}
\label{fig:h=0}
\end{figure}

If $\theta_0\neq 0$ the minimal surface is planar and consists in symmetric difference ${\cal S}_+\Delta{\cal S}_-$ of the surface of the two frames (see Fig. \ref{fig:h=0} in the elliptic case) and the force is vanishing.

In the case of elliptic frame it is easy to calculate the area of the symmetric difference that is proportional to the energy of the film
\begin{eqnarray}
\nonumber
{\cal E}\left(\theta_0\right)=&& 2\gamma\frac{4 R_{\rm m}^2}{\sqrt{1-e^2}} \left[\arctan{\left(\frac{\cot{\frac{\theta_0}{2}}}{\sqrt{1 - e^2}}\right)}\right.\\ 
&&\left.+\arctan{\left(\frac{\tan{\frac{\theta_0}{2}}}{\sqrt{1 - e^2}}\right)} - \frac{\pi}{2}\right].
\end{eqnarray}
By deriving this relation with respect to $\theta_0$ we can obtain the torque
\begin{equation}
\Gamma_z\left(\theta_0\right)=2\gamma \frac{8R_{\rm m}^2 e^2\cos{\left(\theta_0\right)}}{(e^2-2)^2 - e^4\cos^2{\left(\theta_0\right)}}.
\end{equation}

\subsection{Maximal force}
\label{forcemax}

We discuss here the existence of a maximum of the normal force. We recall that the normal force, as discussed in the paper in Section \ref{force:theo}, is directly proportional to the perimeter of ${\cal C}_{\parallel}$, the projection of the neck contour ${\cal C}_{\rm neck}$ in the mid-plane between the frames. Let us first consider the axisymmetric case of an asymmetric catenoid supported by two circular frames of different radii ($R_1 < R_2$) \cite{Salkin}. We can distinguish two cases: (i) if the ratio $R_1/R_2 \leq 0.55$ ($R_1$ is smaller or equal than the critical neck radius) then the neck contour is always virtual (in the prolongation of the minimal surface) and the force is always increasing monotonously up to the critical point; (ii) if $R_1/R_2 > 0.55$, the force increases with $h$ until the neck contour is merging with the smaller frame. At that height, the force is maximum, and as the separation is further increased, the force decreases. We guess that in the more general case of minimal surfaces supported by arbitrary planar frames, if the neck contour is completely inside the surface, then the force is decreasing with height.

\begin{figure}[h!]
\centering
\includegraphics[width=0.75\textwidth]{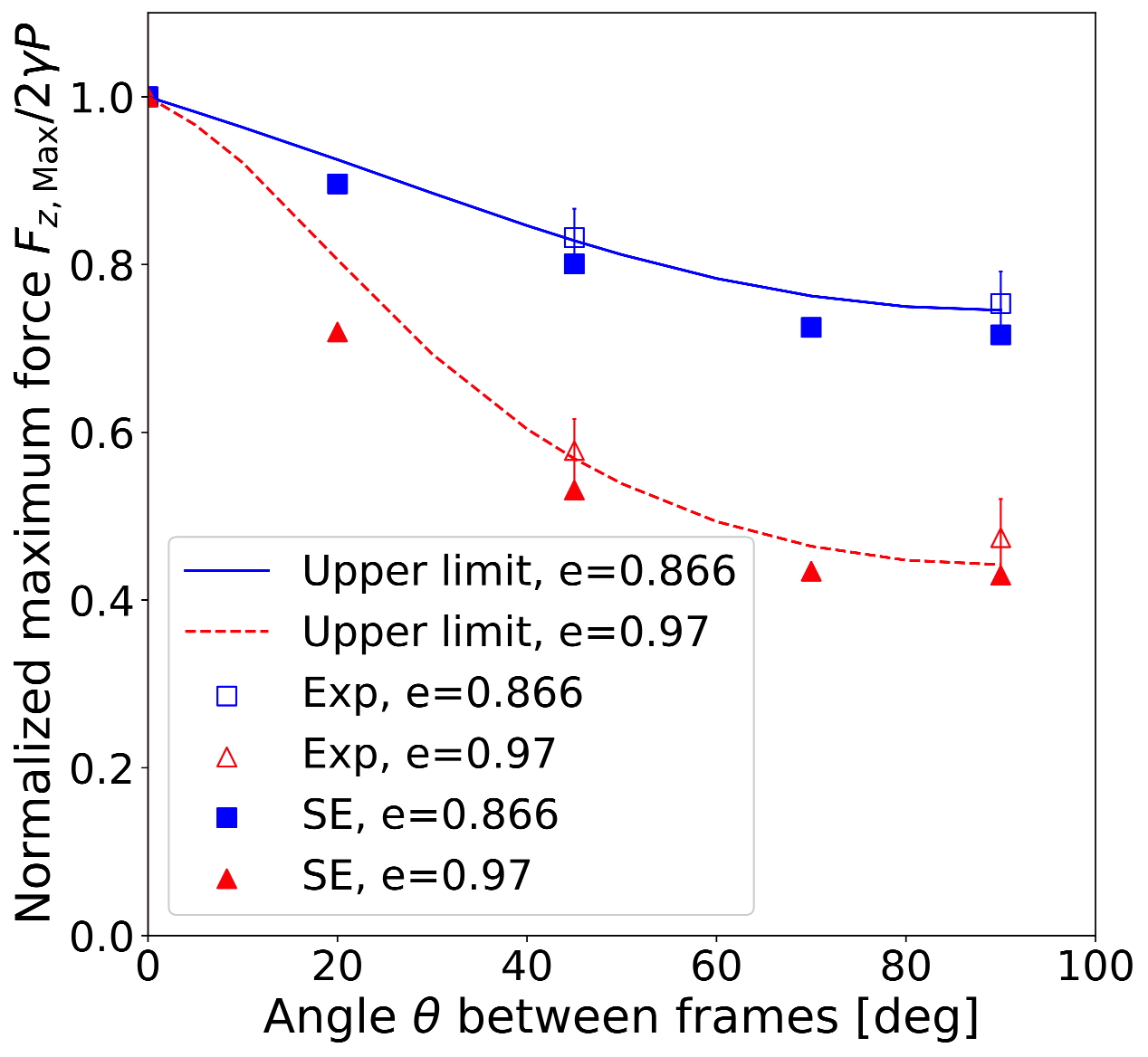}  \caption{Normalized maximum force in the case of elliptic frames ($e=0.866$ in blue and $e=0.97$ in red) obtained experimentally and numerically (SE). Lines correspond to the perimeter ABCD of the grey area on Fig. \ref{fig:h=0} normalized by the perimeter of the frame.}
\label{fig:forcemax}
\end{figure}

In the case of an asymmetric catenoid, when $R_1/R_2 > 0.55$, the maximum value of the force is proportional to the perimeter of the smaller frame, which is also the perimeter of the intersection of the surfaces bounded by the two frames. In the general case, the perimeter of ${\cal C}_{\parallel}$ is upper bounded by the perimeter of the intersection of the projection of the two frames in the mid-plane (grey area on Fig. \ref{fig:h=0}) (It can be proven by integrating the surface tension force along the closed contour ($A_+\rightarrow B_+\rightarrow B_- \rightarrow C_-\rightarrow C_+\rightarrow D_+\rightarrow D_- \rightarrow A_-\rightarrow A_+$, see Fig.  \ref{fig:h=0}). By construction, the parts of the contours connecting the two frames (typically $X_+\rightarrow X_-$ with $X=A, B, C$ or $D$) do not give any contribution to the vertical force, whereas the frame parts give a total vertical force which is lower than the perimeter times $2\gamma$, because the local angle can only give lower values. As the construction of  ${\cal C}_{\parallel}$ ensures that its perimeter is equal to $F_z/(2\gamma)$ it follows that the perimeter of the projected neck contour is shorter than the projected intersection perimeter. This is illustrated in the case of the elliptic frames on Fig. \ref{fig:forcemax} which clearly shows that this theoretical upper bound gives an excellent approximation of the numerical and experimental values of the force.

\pagebreak

\end{document}